\documentclass[preprint,12pt, authoryear]{elsarticle}

\usepackage{amssymb}
\usepackage{deluxetable}
\usepackage{adjustbox}
\journal{Icarus}

\begin{document}

\begin{frontmatter}

%% Title, authors and addresses

%% use the tnoteref command within \title for footnotes;
%% use the tnotetext command for theassociated footnote;
%% use the fnref command within \author or \address for footnotes;
%% use the fntext command for theassociated footnote;
%% use the corref command within \author for corresponding author footnotes;
%% use the cortext command for theassociated footnote;
%% use the ead command for the email address,
%% and the form \ead[url] for the home page:
%% \title{Title\tnoteref{label1}}
%% \tnotetext[label1]{}
%% \author{Name\corref{cor1}\fnref{label2}}
%% \ead{email address}
%% \ead[url]{home page}
%% \fntext[label2]{}
%% \cortext[cor1]{}
%% \address{Address\fnref{label3}}
%% \fntext[label3]{}

\title{A Search for Subkilometer-sized Ordinary Chondrite Like Asteroids in the Main-Belt}

%% use optional labels to link authors explicitly to addresses:
%% \author[label1,label2]{}
%% \address[label1]{}
%% \address[label2]{}

\author[1]{H. W. Lin}
\author[2]{Fumi Yoshida}
\author[3]{Y. T. Chen}
\author[1,4]{W. H. Ip}
\author[1]{C. K. Chang}

\address[1]{Institute of Astronomy, National Central University, Taoyuan 32001, Taiwan}
\address[2]{National Astronomical Observatory of Japan, 2-21-1 Osawa, Mitaka, Tokyo 181-8588, JAPAN}
\address[3]{Institute of Astronomy and Astrophysics, Academia Sinica, P. O. Box 23-141, Taipei 106, Taiwan}
\address[4]{Space Science Institute, Macau University of Science and Technology, Taipa, Macau}

\begin{abstract}
%% Text of abstract

The size-dependent effects of asteroids on surface regolith and collisional lifetimes suggest that small asteroids are younger than large asteroids. 
In this study, we performed multicolor main-belt asteroid (MBA) survey by Subaru telescope/Suprime-Cam to search for subkilometer-sized ordinary chondrite (Q-type) like MBAs. The total survey area was 1.5 deg$^2$ near ecliptic plane and close to the opposition. We detected 150 MBAs with 4 bands ($B$, $V$, $R$, $I$) in this survey. The range of absolute magnitude of detected asteroids was between 13 and 22 magnitude, which is equivalent to the size range of kilometer to sub-kilometer diameter in MBAs.  

From this observation, 75 of 150 MBAs with color uncertainty less than 0.1 were used in the spectral type analysis, and two possible Q-type asteroids were detected. This mean that the Q-type to S-type ratio in MBAs is $<$ 0.05. Meanwhile, the Q/S ratio in near Earth asteroids (NEAs) has been estimated to be 0.5 to 2 \citep{bin04, dan03}. Therefore, Q-type NEAs might be delivered from the main belt region with weathered, S-type surface into near Earth region and then obtain their Q-type, non-weathered surface after undergoing re-surfacing process there. The resurfacing mechanisms could be: 1. dispersal of surface material by tidal effect during planetary encounters \citep{bin10, nes10}, 2. the YORP spin-up induced rotational-fission \citep{pol14} or surface re-arrangement, or 3. thermal degradation \citep{del14}.

\end{abstract}

\begin{keyword}
%% keywords here, in the form: keyword \sep keyword

%% PACS codes here, in the form: \PACS code \sep code

%% MSC codes here, in the form: \MSC code \sep code
%% or \MSC[2008] code \sep code (2000 is the default)

Asteroids, surfaces \sep Asteroids, composition \sep Asteroids \sep Regoliths

\end{keyword}

\end{frontmatter}

%% \linenumbers

%% main text
\section{Introduction}

The taxonomic type of asteroid has been studied extensively to understand the mineral composition of asteroids. It is mostly based on the asteroid's colors and spectra in optical wavelength. Numerous types (such as S, C, D, B and V) have been identified \citep{bow78, tho84, zel85, bus02a, bus02b, dem09, dem13, dem14a}. Space weathering effects and related color-spectrum correlations for the main-belt asteroids (MBAs) and near-Earth asteroids (NEAs) have also been studied \citep{cha96, bin01,  cha04, cla01, jed04,  nes05, wil08, wil10, wil11, tho12}. These studies have been primarily based on relatively larger asteroids; only few studies done for kilometer to sub-kilometer asteroids because of the requirement of large telescopes to determine their colors or spectra, have been conducted.

Asteroids with sizes below the kilometer range are most likely collisional fragments of large asteroids \citep{dav02, mor09}. Therefore, their surfaces should have lower degree of space weathering compared with larger asteroids, which have survived throughout the history of the solar system \citep{bin01, bin04, bus02b}.
Some small, several hundred meter sized NEAs observed in detail while during close approach to the Earth, showed Q-type spectra which are similar to those of the ordinary chondrite (OC) with low degree of space weathering \citep{tho84}. Researchers also reported that a spectral transition could occur between S-type and Q-type asteroids \citep{bin96, bin04, dan03}. These results indicated that S-type asteroids are likely Q-type asteroids, with their surface materials originally characterized by OC-like spectra, but modified by space weathering to the present-day darker and redder spectra. Laboratory experiments \citep{sas01, bru06}) and observations conducted by the NEAR and Hayabusa space missions \citep{cla02,ish07} ) supported this theory.

A large number of Q-type asteroids have been detected in the near-Earth region \citep{bin01, dan03, stu04, dem13}, the ratio of Q/S in NEAs is 0.5 to 2. If Q-type NEAs were produced by collisions, we should also detect Q-type MBAs because the collisional rate in main-belt is higher than that in near-Earth region. While, Q-type asteroids were missing in the main-belt in the earlier studies \citep{bus02a, bus02b, laz04}, more recent observations have detected several Q-type MBAs in the extremely young asteroid family ``Datura dynamical cluster'' \citep{mot08} and the older Koronis family \citep{riv11, tho11}. \citet{car10} classed 3296 of 62576 asteroids as Q-type-like objects in SDSS Moving Object Catalog (SDSSMOC4). \citet{pol14} also detect two Q-type asteroids, (19289) 1996 HY12 and (54827) 2001 NQ8, in the unbound asteroid pairs. 
These results show that Q-type taxonomy is not limited to the NEA population. However, the abundance of Q-type asteroids in the main-belt is still low comparing with that in the near-Earth region. A simple explanation for this low ratio of Q/S in MBAs is that the survey of small MBAs is incomplete, many of Q-type MBAs might be discovered if the observations are able to detect the sub-kilometer-sized MBAs.

The collisional formation model of Q-type NEAs (hereafter, ``standard model'') could be challenged by the rapid process of space weathering effect with solar wind implantation \citep{hap01, ver09}, which timescale could be as short as $\sim 10^4$ to $10^6$ years. Two arguments have been presented.
First, kilometer-sized or large asteroids with collisional lifetimes exceeding $10^8$ years \citep{bot93, bot94} should not display Q-type spectra under long-term space weathering effect. 
However, the existence of kilometer-sized Q-type NEAs, such as (1862) Apollo \citep{stu04}, contradicts the prediction of the ``standard model''.
Second, the high collisional rate in the main-belt region can produce asteroids with fresh surfaces more efficiently. On the other hand, the time scale of transport processes, such as the Yarkovsky effect and small resonances, that insert collisional fragments into the planet-crossing space also exceeds $10^6$ years \citep{rab97, mor03, mig98, bot02, bin04}. Therefore, Q-type NEAs should not be present if they were primarily transferred from the main-belt with a Q-type spectra. This contradicts the observations made in near-Earth space.

An alternative scenario involves a possibility that the surfaces of Q-type NEAs have been reset during planetary encounters, from which the surface materials were removed \citep{nes05} or re-arranged \citep{bin10} by tidal effect. This hypothesis has been tested by several theoretical and observational studies. For example, \citet{mar06} determined that the spectral slope of Q-type asteroids is correlated with planet-crossing frequency. \citet{bin10} and \citet{nes10} suggested that the Q-type NEAs have experienced encounters with the Earth, Venus and then the tidal forces from these terrestrial planets could refresh the asteroidal surfaces. \citet{dem14b} proposed the possibility that this mechanism might also be valid for Mars. As a corollary, the planetary encounter models predicts that Q-type asteroids are rare among MBAs because of the low planetary encounter rate in the main-belt.

Nevertheless, the timescale of space weathering on asteroid surface is still in debate. \citet{wil11} studied 95 asteroids for which span a size and age range of about 1-20 km and 100-3000 Myr, respectively, and measured a space weathering time of $2 \times10^9$ years. This is much longer than the result of fast space weathering \citep{hap01, ver09}. \citet{pol14} also detected a Q-type asteroid in main-belt with age $\sim 10^6$ years indicating that the space weathering timescale should be no less than $\sim 10^6$ years.

The other mechanisms to create Q-type asteroids are correlated with fast rotation and YORP effect: 1. rotational-fission results in the exposure of material from the covered surface of parent asteroid \citep{pol14}, and 2. rotational re-arrangement of asteroid surface material via landslips \citep{sch15, wal12} and partial removal of weathered regolith. These two mechanisms are able to uncover non-weathered materials and display the fresh Q-type spectra. Since the rotation of the smaller astroids are easier to be accelerated by YORP effect, we expect to detect more small size Q-type MBAs if rotational effects are the dominant mechanism of the Q-type asteroid formation.

\citet{del14} reported recently that thermal degradation induced by diurnal temperature variation is able to break up rocks on the asteroid surface rapidly into new regolith layer. They also suggested that asteroids with large diurnal temperature difference (i.e., NEAs) can be cover by fresh regolith characterized by the Q-type spectra. This scenario predicts that more Q-type asteroids should be detected in near Earth space than in main-belt because of the larger diurnal temperature variation of NEAs. Note that the regolith formation by thermal fragmentation does not depend on asteroid size; it may also imply that there is no color-size relation in the NEA population.

From the discussion above, the multicolor observation of kilometer to sub-kilometer diameter MBAs becomes critical to understand the space weathering on S-complex asteroid surface and the formation of Q-type asteroids. We should detect a comparable or even higher fraction of Q-type asteroids in the main-belt than the near-Earth region, if the space weathering timescale is $> 10^8$ years, and the collisional  ``standard model'' dominate the formation of Q-type asteroids. By contrast, if the Q-type asteroid fraction in the main-belt is very low, the Q-type NEAs must form in-situ and other mechanism like the planetary encounter models, rotational-fission/re-arrangement or thermal degradation should be responsible of the formation of Q-type NEAs.

\section{Observations and Data Reduction}
To find sub-kilometer asteroids in the main-belt, we used the data taken by Subaru telescope with Suprime-Cam, which is a prime focus camera with a wide field of view (34' x 27') that consists of 10 CCD chips \citep{miy02}. The observational dates were August 9 and 10, 2004 (UT). 
 
Three fields were surveyed each night for approximately 3.5 hours from the midnight of Hawaii. The seeing size was 0.54-0.70 arcsecond on the first night and 0.77-1.06 arcsecond on the second night. The observational fields were near opposition and close to the ecliptic plane. The center of the coordinates of each observed field is listed in Table~\ref{tab1}.

The images were obtained using four broadband filters: $B$, $V$, $R$ and $I$. The exposure times were 120 sec for the $B$-, $V$- and $R$-bands and 180 sec for the $I$-band. The observations followed the color sequence $RR$-$BB$-$VV$-$RR$-$II$-$RR$ for each fields. The time interval of the first $R$-band set and the second $R$-band set was approximately 80 min. The interval between the second $R$-band set and the third $R$-band set was approximately 60 min. We used this three sets of $R$-band observations to interpolate the R magnitude in the epoch of $B,V$ and $I$-band observations to avoid possible color uncertainty due to the asteroid rotational effect. Detailed description of photometric calibration can be found in Sections 2.2 and 2.3 in detail.
%The observational log is summarized in Table~\ref{tab2}.

\clearpage
 
\begin{table}
\caption{Coordinates of the surveyed fields \label{tab1}}
\begin{tabular}{ccccc}

Aug. 09, 2004\\
\hline
\hline
Field ID & RA & DEC & $\lambda$ & $\beta$ \\
\hline
F1-1 & 21:18:24 & -15:11:00 & 317.306 & 0.488\\
F2-1 & 21:18:24 & -15:41:00 & 317.154 & 0.011\\
F3-1 & 21:18:24 & -16:11:00 & 317.003 & -0.465\\
\hline
 \\
Aug. 10, 2004\\
\hline
\hline
Field ID & RA & DEC & $\lambda$ & $\beta$ \\
\hline
F1-2 & 21:22:12 & -14:54:00 & 318.266 & 0.478\\
F2-2 & 21:22:12 & -15:24:00 & 318.112 & 0.002\\
F3-2 & 21:22:12 & -15:54:00 & 317.959 & -0.474\\
\hline
%\tablecomments{$\lambda$ and $\beta$ are ecliptic longitude and latitude.}
\end{tabular}
\end{table}

\subsection{Detection of Moving Objects}
To detect moving objects in relatively crowded fields, we first stacked all 12 exposures in each field to obtain deep images. We then used these deep images as the source images to generate the reference stationary catalogs and remove all stationary sources in every exposure.

After removing the stationary sources, we used the KDTree-based nearest neighborhood search method to identify the detection pairs in every two consecutive exposures. The pairs detected in the first set of consecutive exposures were used to determine the main vectors for predicting the possible locations in the other five exposure pairs. We then searched for the corresponding pairs at the predicted locations of the other set of consecutive exposure. Once the entire set of six pairs (i.e., 12 detections in total) was identified, the complete set was passed to the $Orbfit$ code \citep{ber00} to ensure that the orbital solution are reasonable; the semi-major axis was between 2 AU and 5 AU, and the fitting residual was smaller than 0.5''. Under these stringent conditions, all moving objects detected are real and complete color measurements were performed for all of them. The asteroid detection list  is summarized in Table~\ref{tab2}.

\begin{deluxetable}{lccccccccccccccccccccc}
\rotate
\tabletypesize{\scriptsize}
\tablecaption{Asteroid list\label{tab2}}
\tablewidth{0pt}
\setlength{\tabcolsep}{0.03in} 
\tablehead{\colhead{RA ($^\circ$)} & \colhead{DEC ($^\circ$)} & \colhead{Epoch (MJD)} & \colhead{a (AU)$^a$} & \colhead{i ($^\circ$)$^a$} & \colhead{H$_V$} & \colhead{B} & \colhead{B$_{err}$ } & \colhead{V} & \colhead{V$_{err}$}& \colhead{R} & \colhead{R$_{err}$}& \colhead{I} & \colhead{I$_{err}$} &A$^b$ & A$_{err}^b$& \colhead{Comment} & \colhead{a (AU)$^c$} & \colhead{e $^c$} & \colhead{i ($^\circ)^c$} }
\startdata
319.48272	&	-15.40419	&	53226.529043	&	2.84	&	2.31	&	18.103	&	22.561	&	0.014	&	21.690	&	0.002	&	21.277	&	0.011	&	20.926	&	0.008	&	0.088	&	0.011	&		&		&		&		\\
319.39225	&	-15.22197	&	53226.529043	&	2.90	&	4.69	&	19.612	&	24.146	&	0.021	&	23.320	&	0.002	&	22.981	&	0.017	&	22.584	&	0.001	&	0.004	&	0.106	&		&		&		&		\\
319.69678	&	-15.20883	&	53226.529043	&	3.21	&	0.03	&	18.980	&	23.970	&	0.025	&	23.231	&	0.037	&	22.980	&	0.038	&	22.661	&	0.001	&	-0.119	&	0.053	&		&		&		&		\\
319.46102	&	-15.20154	&	53226.529043	&	3.09	&	16.65	&	19.088	&	23.911	&	0.008	&	23.141	&	0.003	&	22.749	&	0.017	&	22.246	&	0.016	&	0.002	&	0.034	&		&		&		&		\\
319.46236	&	-15.19466	&	53226.529043	&	2.25	&	5.50	&	18.683	&	21.583	&	0.030	&	20.930	&	0.012	&	20.607	&	0.014	&	20.563	&	0.027	&	-0.130	&	0.024	&		&		&		&		\\
319.58996	&	-15.18535	&	53226.529043	&	2.55	&	8.45	&	19.353	&	23.149	&	0.001	&	22.336	&	0.026	&	21.922	&	0.013	&	21.560	&	0.012	&	0.048	&	0.048	&		&		&		&		\\
319.57579	&	-15.16933	&	53226.529043	&	2.84	&	9.71	&	18.900	&	23.027	&	0.012	&	22.493	&	0.006	&	21.859	&	0.012	&	21.519	&	0.023	&	0.006	&	0.020	&		&		&		&		\\
319.38260	&	-15.16018	&	53226.529043	&	3.05	&	0.16	&	19.211	&	23.882	&	0.087	&	23.200	&	0.015	&	23.056	&	0.032	&	22.667	&	0.020	&	-0.236	&	0.140	&		&		&		&		\\
319.34533	&	-15.16815	&	53226.529043	&	2.56	&	23.00	&	21.014	&	25.138	&	0.016	&	24.025	&	0.023	&	23.339	&	0.240	&	22.887	&	0.050	&	0.452	&	0.351	&		&		&		&		\\
319.63049	&	-15.14830	&	53226.529043	&	2.92	&	1.07	&	20.022	&	24.513	&	0.255	&	23.774	&	0.046	&	23.602	&	0.048	&	23.087	&	0.125	&	-0.176	&	0.245	&		&		&		&		\\
319.86432	&	-15.11508	&	53226.529043	&	2.53	&	11.95	&	15.351	&	19.177	&	0.002	&	18.279	&	0.003	&	17.823	&	0.003	&	17.423	&	0.009	&	0.138	&	0.003	&		&		&		&		\\
319.37775	&	-15.09740	&	53226.529043	&	2.85	&	2.08	&	19.851	&	24.105	&	0.030	&	23.461	&	0.012	&	23.022	&	0.064	&	22.618	&	0.041	&	-0.054	&	0.050	&		&		&		&		\\
319.82910	&	-15.09154	&	53226.529043	&	3.46	&	0.68	&	18.449	&	23.810	&	0.041	&	23.106	&	0.028	&	22.772	&	0.037	&	22.467	&	0.009	&	-0.086	&	0.044	&		&		&		&		\\
319.58372	&	-15.09950	&	53226.529043	&	2.61	&	19.12	&	18.133	&	22.075	&	0.010	&	21.258	&	0.003	&	20.762	&	0.013	&	20.367	&	0.004	&	0.109	&	0.008	&		&		&		&		\\
319.81048	&	-15.08243	&	53226.529043	&	2.99	&	13.25	&	19.227	&	23.837	&	0.042	&	23.096	&	0.028	&	22.738	&	0.049	&	22.321	&	0.026	&	-0.042	&	0.040	&		&		&		&		\\
319.72990	&	-15.05428	&	53226.529043	&	2.91	&	0.46	&	19.240	&	23.795	&	0.052	&	22.967	&	0.004	&	22.727	&	0.051	&	22.365	&	0.019	&	-0.065	&	0.104	&		&		&		&		\\
319.31365	&	-15.03409	&	53226.529043	&	2.71	&	7.15	&	19.777	&	23.905	&	0.011	&	23.108	&	0.006	&	22.726	&	0.015	&	22.375	&	0.057	&	0.014	&	0.045	&		&		&		&		\\
319.86975	&	-15.03411	&	53226.529043	&	2.91	&	4.80	&	15.174	&	19.783	&	0.008	&	18.904	&	0.000	&	18.410	&	0.003	&	17.941	&	0.002	&	0.151	&	0.008	&		&		&		&		\\
319.32668	&	-15.01941	&	53226.529043	&	2.92	&	1.91	&	17.177	&	21.680	&	0.006	&	20.918	&	0.002	&	20.463	&	0.007	&	20.018	&	0.020	&	0.041	&	0.006	&		&		&		&		\\
319.70512	&	-14.98392	&	53226.529043	&	2.29	&	5.76	&	19.579	&	22.853	&	0.033	&	21.928	&	0.013	&	21.657	&	0.003	&	21.035	&	0.024	&	0.025	&	0.031	&		&		&		&		\\
319.76720	&	-15.30467	&	53226.529043	&	3.28	&	10.14	&	17.388	&	22.444	&	0.001	&	21.760	&	0.002	&	21.409	&	0.008	&	21.071	&	0.003	&	-0.089	&	0.027	&		&		&		&		\\
319.88038	&	-15.29547	&	53226.529043	&	2.41	&	2.13	&	19.033	&	22.554	&	0.020	&	21.687	&	0.012	&	21.228	&	0.008	&	20.831	&	0.008	&	0.118	&	0.019	&		&		&		&		\\
319.74818	&	-15.28862	&	53226.529043	&	2.68	&	4.36	&	15.391	&	19.505	&	0.001	&	18.654	&	0.003	&	18.189	&	0.001	&	17.780	&	0.020	&	0.111	&	0.003	&	(62523) 2000 SW24	&	2.7	&	0.07	&	4.28	\\
319.30847	&	-15.28169	&	53226.529043	&	3.16	&	10.69	&	18.476	&	23.424	&	0.028	&	22.650	&	0.021	&	22.377	&	0.022	&	21.998	&	0.011	&	-0.080	&	0.041	&		&		&		&		\\
319.86194	&	-15.28194	&	53226.529043	&	2.97	&	8.94	&	19.373	&	23.631	&	0.085	&	23.205	&	0.163	&	22.552	&	0.029	&	22.227	&	0.050	&	-0.057	&	0.208	&		&		&		&		\\
319.71990	&	-15.27244	&	53226.529043	&	3.03	&	14.48	&	20.092	&	25.335	&	0.035	&	24.039	&	0.074	&	23.739	&	0.039	&	23.551	&	0.052	&	0.308	&	0.075	&		&		&		&		\\
319.35283	&	-15.27298	&	53226.529043	&	2.72	&	4.47	&	20.254	&	24.188	&	0.004	&	23.599	&	0.053	&	23.245	&	0.020	&	22.988	&	0.002	&	-0.153	&	0.064	&		&		&		&		\\
319.83410	&	-15.23152	&	53226.529043	&	3.01	&	9.89	&	17.812	&	22.470	&	0.004	&	21.713	&	0.015	&	21.393	&	0.016	&	21.670	&	0.434	&	-0.058	&	0.025	&		&		&		&		\\
319.62572	&	-15.22084	&	53226.529043	&	2.91	&	1.49	&	13.441	&	17.950	&	0.005	&	17.165	&	0.011	&	16.909	&	0.039	&	16.808	&	0.028	&	-0.084	&	0.018	&		&		&		&		\\
319.69906	&	-15.22543	&	53226.529043	&	3.06	&	0.96	&	19.460	&	24.116	&	0.056	&	23.453	&	0.063	&	23.019	&	0.036	&	22.602	&	0.004	&	-0.045	&	0.067	&		&		&		&		\\
320.70343	&	-15.12162	&	53227.533671	&	2.54	&	0.60	&	19.229	&	23.109	&	0.009	&	22.184	&	0.009	&	21.719	&	0.019	&	21.398	&	0.002	&	0.163	&	0.035	&		&		&		&		\\
320.65551	&	-15.05483	&	53227.533671	&	2.99	&	1.51	&	16.549	&	21.252	&	0.001	&	20.414	&	0.004	&	19.955	&	0.004	&	19.528	&	0.005	&	0.097	&	0.004	&		&		&		&		\\
320.68576	&	-15.01347	&	53227.533671	&	2.88	&	3.55	&	18.366	&	22.788	&	0.006	&	22.040	&	0.021	&	21.655	&	0.020	&	21.333	&	0.004	&	-0.019	&	0.016	&		&		&		&		\\
320.59647	&	-15.00993	&	53227.533671	&	3.03	&	10.02	&	19.492	&	24.327	&	0.030	&	23.437	&	0.034	&	23.172	&	0.025	&	22.874	&	0.016	&	-0.003	&	0.047	&		&		&		&		\\
320.33657	&	-14.97422	&	53227.533671	&	3.04	&	2.47	&	17.160	&	22.045	&	0.004	&	21.123	&	0.012	&	20.678	&	0.007	&	20.248	&	0.002	&	0.146	&	0.015	&	(252718) 2002 CQ194	&	3.01	&	0.08	&	3.47	\\
320.66292	&	-14.89674	&	53227.533671	&	3.05	&	0.55	&	19.167	&	24.113	&	0.025	&	23.148	&	0.110	&	22.830	&	0.077	&	22.386	&	0.042	&	0.087	&	0.092	&		&		&		&		\\
320.44262	&	-14.85440	&	53227.533671	&	2.64	&	3.27	&	19.728	&	24.209	&	0.063	&	22.907	&	0.191	&	22.512	&	0.049	&	22.306	&	0.037	&	0.380	&	0.220	&		&		&		&		\\
320.58374	&	-14.84279	&	53227.533671	&	3.12	&	0.43	&	16.584	&	21.393	&	0.009	&	20.680	&	0.009	&	20.331	&	0.005	&	19.975	&	0.001	&	-0.069	&	0.016	&		&		&		&		\\
320.26807	&	-14.82940	&	53227.533671	&	2.88	&	2.04	&	16.242	&	20.761	&	0.006	&	19.912	&	0.001	&	19.504	&	0.005	&	19.183	&	0.003	&	0.069	&	0.015	&	Q-type candidate	&		&		&		\\
320.54242	&	-14.80967	&	53227.533671	&	2.94	&	9.75	&	19.503	&	23.711	&	0.160	&	23.289	&	0.049	&	22.955	&	0.029	&	22.512	&	0.023	&	-0.285	&	0.152	&		&		&		&		\\
320.52186	&	-14.78980	&	53227.533671	&	3.11	&	2.37	&	18.405	&	23.366	&	0.013	&	22.495	&	0.016	&	21.906	&	0.078	&	21.411	&	0.022	&	0.213	&	0.031	&		&		&		&		\\
319.51576	&	-15.68548	&	53226.534619	&	2.50	&	9.54	&	20.173	&	23.905	&	0.032	&	23.036	&	0.001	&	22.675	&	0.011	&	22.290	&	0.010	&	0.050	&	0.033	&		&		&		&		\\
319.85629	&	-15.67499	&	53226.534619	&	2.57	&	6.24	&	20.294	&	24.496	&	0.009	&	23.330	&	0.062	&	23.168	&	0.013	&	22.646	&	0.090	&	0.119	&	0.062	&		&		&		&		\\
319.75825	&	-15.67583	&	53226.534619	&	2.31	&	0.80	&	21.302	&	24.414	&	0.004	&	23.709	&	0.069	&	23.215	&	0.022	&	22.703	&	0.048	&	0.028	&	0.108	&		&		&		&		\\
319.67486	&	-15.66787	&	53226.534619	&	2.14	&	1.03	&	22.288	&	24.940	&	0.189	&	24.237	&	0.004	&	23.481	&	0.063	&	23.408	&	0.000	&	0.212	&	0.158	&		&		&		&		\\
319.67821	&	-15.66154	&	53226.534619	&	3.23	&	2.79	&	19.251	&	23.904	&	0.309	&	23.531	&	0.024	&	23.224	&	0.036	&	22.888	&	0.035	&	-0.339	&	0.221	&		&		&		&		\\
319.38817	&	-15.64510	&	53226.534619	&	2.68	&	8.59	&	20.630	&	24.276	&	0.162	&	23.901	&	0.102	&	23.324	&	0.039	&	23.055	&	0.008	&	-0.146	&	0.141	&		&		&		&		\\
319.50101	&	-15.63361	&	53226.534619	&	2.41	&	3.41	&	14.861	&	18.618	&	0.007	&	17.524	&	0.012	&	17.207	&	0.014	&	16.954	&	0.012	&	0.178	&	0.013	&		&		&		&		\\
319.70150	&	-15.63029	&	53226.534619	&	2.19	&	7.19	&	16.856	&	19.749	&	0.004	&	18.932	&	0.003	&	18.346	&	0.043	&	18.092	&	0.003	&	0.172	&	0.005	&	(204290) 2004 PV21	&	2.17	&	0.1	&	6.36	\\
319.72054	&	-15.60861	&	53226.534619	&	2.60	&	3.43	&	17.202	&	21.007	&	0.002	&	20.289	&	0.006	&	19.983	&	0.006	&	19.638	&	0.000	&	-0.095	&	0.005	&		&		&		&		\\
319.36405	&	-15.57624	&	53226.534619	&	2.92	&	7.74	&	18.437	&	22.750	&	0.010	&	22.180	&	0.111	&	21.635	&	0.008	&	21.247	&	0.008	&	-0.031	&	0.106	&		&		&		&		\\
319.30469	&	-15.51043	&	53226.534619	&	2.93	&	1.91	&	19.062	&	23.533	&	0.026	&	22.831	&	0.012	&	22.508	&	0.013	&	22.208	&	0.022	&	-0.095	&	0.063	&		&		&		&		\\
319.30965	&	-15.50791	&	53226.534619	&	2.52	&	3.71	&	21.024	&	24.562	&	0.079	&	23.941	&	0.050	&	23.454	&	0.096	&	23.684	&	0.008	&	-0.037	&	0.143	&		&		&		&		\\
319.36096	&	-15.50186	&	53226.534619	&	2.54	&	17.38	&	19.236	&	22.934	&	0.044	&	22.198	&	0.017	&	21.540	&	0.014	&	21.088	&	0.000	&	0.166	&	0.040	&		&		&		&		\\
319.54940	&	-15.87029	&	53226.534619	&	2.43	&	6.49	&	19.608	&	23.105	&	0.043	&	22.312	&	0.003	&	21.859	&	0.025	&	21.406	&	0.009	&	0.062	&	0.034	&		&		&		&		\\
319.35954	&	-15.87175	&	53226.534619	&	2.69	&	0.02	&	20.663	&	24.784	&	0.009	&	23.956	&	0.019	&	23.692	&	0.048	&	23.171	&	0.112	&	-0.048	&	0.022	&		&		&		&		\\
319.60794	&	-15.85584	&	53226.534619	&	2.61	&	14.67	&	20.238	&	24.104	&	0.094	&	23.357	&	0.063	&	22.864	&	0.014	&	22.486	&	0.017	&	0.057	&	0.101	&		&		&		&		\\
319.84755	&	-15.86667	&	53226.534619	&	3.00	&	13.01	&	18.785	&	23.621	&	0.263	&	22.670	&	0.008	&	22.254	&	0.023	&	21.901	&	0.009	&	0.147	&	0.187	&		&		&		&		\\
319.80565	&	-15.84336	&	53226.534619	&	3.13	&	8.43	&	15.163	&	20.077	&	0.000	&	19.285	&	0.011	&	18.832	&	0.007	&	18.504	&	0.002	&	0.061	&	0.012	&		&		&		&		\\
319.74746	&	-15.84407	&	53226.534619	&	2.65	&	0.94	&	19.940	&	23.945	&	0.002	&	23.142	&	0.029	&	22.692	&	0.020	&	22.297	&	0.038	&	0.066	&	0.027	&		&		&		&		\\
319.49048	&	-15.83221	&	53226.534619	&	2.84	&	2.87	&	17.144	&	21.536	&	0.010	&	20.732	&	0.007	&	20.299	&	0.004	&	19.942	&	0.002	&	0.055	&	0.009	&		&		&		&		\\
319.61032	&	-15.82195	&	53226.534619	&	2.64	&	2.77	&	20.579	&	24.493	&	0.038	&	23.763	&	0.031	&	23.312	&	0.033	&	22.931	&	0.013	&	0.015	&	0.060	&		&		&		&		\\
319.46590	&	-15.82367	&	53226.534619	&	3.11	&	1.35	&	17.336	&	21.778	&	0.003	&	21.417	&	0.001	&	20.367	&	0.399	&	20.501	&	0.008	&	0.177	&	0.009	&		&		&		&		\\
319.70470	&	-15.80420	&	53226.534619	&	2.43	&	6.59	&	20.853	&	24.404	&	0.020	&	23.553	&	0.026	&	23.061	&	0.040	&	22.749	&	0.013	&	0.130	&	0.029	&		&		&		&		\\
319.47202	&	-15.79704	&	53226.534619	&	2.77	&	5.62	&	20.608	&	24.561	&	0.061	&	24.058	&	0.024	&	23.762	&	0.068	&	23.680	&	0.044	&	-0.255	&	0.061	&		&		&		&		\\
319.59842	&	-15.79135	&	53226.534619	&	2.77	&	3.68	&	19.621	&	23.884	&	0.015	&	23.078	&	0.008	&	22.683	&	0.032	&	22.294	&	0.047	&	0.030	&	0.048	&		&		&		&		\\
319.61758	&	-15.79291	&	53226.534619	&	2.23	&	4.60	&	20.205	&	22.574	&	0.211	&	22.403	&	0.213	&	21.716	&	0.010	&	21.393	&	0.017	&	-0.213	&	0.232	&		&		&		&		\\
319.45546	&	-15.78328	&	53226.534619	&	2.49	&	4.66	&	17.988	&	21.554	&	0.000	&	20.831	&	0.004	&	20.460	&	0.005	&	20.094	&	0.001	&	-0.047	&	0.005	&		&		&		&		\\
319.75107	&	-15.77375	&	53226.534619	&	2.71	&	3.71	&	20.252	&	24.560	&	0.019	&	23.586	&	0.039	&	23.325	&	0.037	&	22.823	&	0.042	&	0.053	&	0.048	&		&		&		&		\\
319.71019	&	-15.75217	&	53226.534619	&	2.47	&	1.68	&	19.898	&	23.581	&	0.031	&	22.691	&	0.014	&	22.219	&	0.011	&	21.775	&	0.044	&	0.143	&	0.042	&		&		&		&		\\
319.39739	&	-15.74895	&	53226.534619	&	2.77	&	6.53	&	18.324	&	22.572	&	0.030	&	21.779	&	0.007	&	21.291	&	0.007	&	20.898	&	0.001	&	0.086	&	0.022	&		&		&		&		\\
319.43563	&	-15.73506	&	53226.534619	&	2.64	&	1.59	&	20.248	&	24.317	&	0.005	&	23.435	&	0.014	&	23.026	&	0.017	&	22.388	&	0.156	&	0.093	&	0.037	&		&		&		&		\\
319.35384	&	-15.72188	&	53226.534619	&	2.80	&	5.80	&	18.519	&	22.810	&	0.017	&	22.026	&	0.005	&	21.627	&	0.028	&	21.333	&	0.010	&	0.017	&	0.013	&		&		&		&		\\
320.42609	&	-15.57531	&	53227.539246	&	2.22	&	3.42	&	15.012	&	17.950	&	0.005	&	17.171	&	0.005	&	16.604	&	0.007	&	16.435	&	0.037	&	0.132	&	0.008	&	(6919) Tomonaga	&	2.26	&	0.1	&	5.16	\\
320.79772	&	-15.55019	&	53227.539246	&	2.72	&	4.30	&	18.238	&	22.254	&	0.019	&	21.593	&	0.014	&	21.240	&	0.014	&	20.890	&	0.024	&	-0.103	&	0.024	&		&		&		&		\\
320.71232	&	-15.53839	&	53227.539246	&	3.04	&	2.51	&	18.195	&	22.782	&	0.007	&	22.164	&	0.023	&	21.677	&	0.063	&	21.371	&	0.003	&	-0.039	&	0.029	&		&		&		&		\\
320.82430	&	-15.53347	&	53227.539246	&	2.54	&	3.91	&	19.103	&	22.670	&	0.042	&	22.071	&	0.034	&	21.274	&	0.062	&	20.832	&	0.022	&	0.167	&	0.085	&		&		&		&		\\
320.32744	&	-15.53627	&	53227.539246	&	2.28	&	3.97	&	17.616	&	20.842	&	0.008	&	19.942	&	0.017	&	19.459	&	0.004	&	19.066	&	0.003	&	0.158	&	0.014	&		&		&		&		\\
320.33516	&	-15.52610	&	53227.539246	&	2.38	&	3.91	&	18.111	&	21.565	&	0.002	&	20.702	&	0.018	&	20.258	&	0.005	&	19.823	&	0.014	&	0.104	&	0.013	&		&		&		&		\\
320.67588	&	-15.49924	&	53227.539246	&	3.16	&	10.05	&	18.341	&	23.253	&	0.030	&	22.519	&	0.004	&	22.227	&	0.004	&	21.915	&	0.009	&	-0.094	&	0.034	&		&		&		&		\\
320.41969	&	-15.47944	&	53227.539246	&	2.37	&	7.19	&	20.047	&	23.933	&	0.117	&	22.608	&	0.009	&	22.424	&	0.060	&	22.051	&	0.018	&	0.247	&	0.091	&		&		&		&		\\
320.58908	&	-15.46737	&	53227.539246	&	3.30	&	4.38	&	15.615	&	20.736	&	0.013	&	20.020	&	0.001	&	19.653	&	0.023	&	19.338	&	0.013	&	-0.054	&	0.009	&		&		&		&		\\
320.28343	&	-15.45079	&	53227.539246	&	2.72	&	5.32	&	17.285	&	21.499	&	0.001	&	20.642	&	0.010	&	20.278	&	0.007	&	19.797	&	0.055	&	0.043	&	0.010	&		&		&		&		\\
320.80845	&	-15.38456	&	53227.539246	&	3.00	&	7.79	&	15.556	&	20.261	&	0.008	&	19.456	&	0.002	&	19.141	&	0.005	&	18.844	&	0.005	&	-0.028	&	0.006	&		&		&		&		\\
320.49664	&	-15.38502	&	53227.539246	&	2.89	&	2.26	&	19.593	&	24.077	&	0.132	&	23.271	&	0.006	&	22.923	&	0.216	&	22.194	&	0.003	&	-0.004	&	0.099	&		&		&		&		\\
320.63919	&	-15.37706	&	53227.539246	&	3.30	&	14.74	&	17.682	&	23.284	&	0.057	&	22.088	&	0.092	&	22.118	&	0.010	&	21.781	&	0.005	&	0.004	&	0.090	&		&		&		&		\\
320.63703	&	-15.38200	&	53227.539246	&	2.57	&	10.28	&	18.101	&	22.082	&	0.130	&	21.137	&	0.001	&	20.780	&	0.011	&	20.310	&	0.012	&	0.100	&	0.093	&		&		&		&		\\
320.54423	&	-15.37073	&	53227.539246	&	2.68	&	6.37	&	19.424	&	23.435	&	0.044	&	22.698	&	0.013	&	22.313	&	0.014	&	21.963	&	0.012	&	-0.027	&	0.083	&		&		&		&		\\
320.32559	&	-15.33817	&	53227.539246	&	3.05	&	1.41	&	17.527	&	22.273	&	0.004	&	21.516	&	0.019	&	21.192	&	0.005	&	20.675	&	0.000	&	-0.055	&	0.014	&		&		&		&		\\
320.76977	&	-15.30567	&	53227.539246	&	3.02	&	0.90	&	17.848	&	22.515	&	0.017	&	21.767	&	0.024	&	21.427	&	0.007	&	21.112	&	0.009	&	-0.051	&	0.024	&		&		&		&		\\
320.70791	&	-15.28156	&	53227.539246	&	2.34	&	1.87	&	18.318	&	21.534	&	0.002	&	20.805	&	0.001	&	20.412	&	0.006	&	20.038	&	0.007	&	-0.027	&	0.005	&	(2014) AR8	&	2.4	&	0.14	&	1.72	\\
320.45096	&	-15.27708	&	53227.539246	&	2.24	&	1.83	&	17.172	&	20.377	&	0.005	&	19.389	&	0.008	&	18.891	&	0.010	&	18.554	&	0.021	&	0.231	&	0.009	&	(151733) 2003 BA88	&	2.32	&	0.15	&	2.64	\\
320.64892	&	-15.22799	&	53227.539246	&	2.45	&	2.77	&	19.498	&	23.051	&	0.010	&	22.241	&	0.035	&	21.872	&	0.023	&	21.580	&	0.041	&	0.014	&	0.030	&		&		&		&		\\
320.63783	&	-15.20756	&	53227.539246	&	2.72	&	0.08	&	18.807	&	22.916	&	0.029	&	22.153	&	0.025	&	21.774	&	0.014	&	21.434	&	0.021	&	-0.012	&	0.037	&		&		&		&		\\
319.35421	&	-16.19792	&	53226.540177	&	3.11	&	0.65	&	18.054	&	22.743	&	0.020	&	22.135	&	0.010	&	21.703	&	0.009	&	21.590	&	0.001	&	-0.084	&	0.024	&		&		&		&		\\
319.60419	&	-16.39941	&	53226.540177	&	2.42	&	4.00	&	20.306	&	23.827	&	0.051	&	22.981	&	0.184	&	22.668	&	0.397	&	22.918	&	0.070	&	0.000	&	0.173	&		&		&		&		\\
319.34255	&	-16.38858	&	53226.540177	&	3.78	&	13.11	&	17.136	&	23.051	&	0.021	&	22.249	&	0.008	&	21.799	&	0.021	&	21.476	&	0.024	&	0.065	&	0.021	&	Q-type candidate	&		&		&		\\
319.62069	&	-16.37934	&	53226.540177	&	2.82	&	9.67	&	18.962	&	23.206	&	0.054	&	22.507	&	0.026	&	22.084	&	0.015	&	21.272	&	0.299	&	-0.027	&	0.047	&		&		&		&		\\
319.33731	&	-16.17518	&	53226.540177	&	2.84	&	2.81	&	20.090	&	24.511	&	0.038	&	23.689	&	0.048	&	23.318	&	0.080	&	23.047	&	0.076	&	0.024	&	0.098	&		&		&		&		\\
319.57251	&	-16.37746	&	53226.540177	&	2.62	&	2.94	&	18.422	&	22.469	&	0.004	&	21.555	&	0.006	&	21.091	&	0.008	&	20.694	&	0.008	&	0.155	&	0.015	&		&		&		&		\\
319.86114	&	-16.36801	&	53226.540177	&	2.35	&	1.14	&	20.808	&	24.117	&	0.028	&	23.307	&	0.025	&	22.627	&	0.025	&	22.147	&	0.018	&	0.234	&	0.047	&		&		&		&		\\
319.64018	&	-16.13018	&	53226.540177	&	2.98	&	12.47	&	18.580	&	23.257	&	0.085	&	22.428	&	0.008	&	22.049	&	0.007	&	21.657	&	0.002	&	0.035	&	0.061	&		&		&		&		\\
319.54882	&	-16.12899	&	53226.540177	&	2.77	&	6.70	&	19.475	&	23.585	&	0.022	&	22.919	&	0.019	&	22.515	&	0.041	&	22.737	&	0.081	&	-0.064	&	0.036	&		&		&		&		\\
319.32721	&	-16.11899	&	53226.540177	&	2.95	&	15.29	&	16.878	&	21.484	&	0.003	&	20.680	&	0.007	&	20.161	&	0.011	&	19.713	&	0.002	&	0.115	&	0.014	&		&		&		&		\\
319.60643	&	-16.12420	&	53226.540177	&	2.59	&	3.05	&	18.688	&	22.621	&	0.009	&	21.759	&	0.003	&	21.339	&	0.008	&	20.964	&	0.009	&	0.087	&	0.014	&		&		&		&		\\
319.82640	&	-16.10238	&	53226.540177	&	2.59	&	3.40	&	20.109	&	23.909	&	0.049	&	23.185	&	0.051	&	22.704	&	0.034	&	22.295	&	0.010	&	0.032	&	0.064	&		&		&		&		\\
319.51692	&	-16.09192	&	53226.540177	&	3.02	&	7.10	&	18.453	&	23.231	&	0.004	&	22.387	&	0.012	&	21.858	&	0.017	&	21.456	&	0.011	&	0.150	&	0.025	&		&		&		&		\\
319.83680	&	-16.07527	&	53226.540177	&	2.35	&	2.16	&	19.252	&	22.467	&	0.011	&	21.753	&	0.005	&	21.402	&	0.014	&	21.058	&	0.002	&	-0.067	&	0.011	&		&		&		&		\\
319.76208	&	-16.06889	&	53226.540177	&	2.96	&	11.26	&	18.247	&	22.939	&	0.010	&	22.072	&	0.096	&	21.766	&	0.015	&	21.404	&	0.007	&	0.009	&	0.103	&		&		&		&		\\
319.72250	&	-16.06501	&	53226.540177	&	2.67	&	7.38	&	18.203	&	22.303	&	0.007	&	21.451	&	0.013	&	21.008	&	0.007	&	20.626	&	0.008	&	0.096	&	0.019	&		&		&		&		\\
319.36741	&	-16.05471	&	53226.540177	&	3.06	&	7.98	&	18.246	&	23.006	&	0.009	&	22.250	&	0.034	&	21.893	&	0.009	&	21.517	&	0.008	&	-0.033	&	0.067	&		&		&		&		\\
319.47395	&	-16.04208	&	53226.540177	&	3.07	&	8.63	&	19.673	&	24.456	&	0.005	&	23.684	&	0.012	&	23.343	&	0.016	&	22.974	&	0.008	&	-0.033	&	0.055	&		&		&		&		\\
319.85396	&	-16.35700	&	53226.540177	&	2.91	&	1.61	&	18.683	&	23.172	&	0.046	&	22.411	&	0.039	&	22.125	&	0.032	&	21.742	&	0.005	&	-0.080	&	0.046	&		&		&		&		\\
319.72944	&	-16.02779	&	53226.540177	&	2.52	&	6.36	&	19.079	&	22.897	&	0.021	&	22.000	&	0.007	&	21.660	&	0.026	&	21.253	&	0.010	&	0.055	&	0.024	&		&		&		&		\\
319.73087	&	-16.35587	&	53226.540177	&	2.75	&	1.16	&	16.955	&	21.061	&	0.009	&	20.376	&	0.012	&	19.827	&	0.005	&	19.377	&	0.020	&	0.052	&	0.014	&		&		&		&		\\
319.76130	&	-16.02214	&	53226.540177	&	2.68	&	1.49	&	20.524	&	24.626	&	0.030	&	23.783	&	0.031	&	23.335	&	0.025	&	23.170	&	0.004	&	0.093	&	0.108	&		&		&		&		\\
319.58537	&	-16.01515	&	53226.540177	&	2.65	&	1.79	&	20.121	&	24.249	&	0.022	&	23.332	&	0.076	&	22.790	&	0.057	&	22.347	&	0.099	&	0.212	&	0.083	&		&		&		&		\\
319.61491	&	-16.00706	&	53226.540177	&	2.75	&	4.22	&	19.611	&	23.959	&	0.006	&	23.030	&	0.025	&	22.592	&	0.022	&	22.195	&	0.011	&	0.147	&	0.021	&		&		&		&		\\
319.43173	&	-16.00280	&	53226.540177	&	3.02	&	0.46	&	13.083	&	17.621	&	0.007	&	17.015	&	0.005	&	16.939	&	0.030	&	16.792	&	0.008	&	-0.338	&	0.007	&		&		&		&		\\
319.72263	&	-16.35138	&	53226.540177	&	2.98	&	5.89	&	19.868	&	24.295	&	0.106	&	23.729	&	0.042	&	23.196	&	0.029	&	22.892	&	0.087	&	-0.043	&	0.107	&		&		&		&		\\
319.62979	&	-15.96643	&	53226.540177	&	2.81	&	2.03	&	17.485	&	21.699	&	0.010	&	21.014	&	0.007	&	20.651	&	0.617	&	20.661	&	0.320	&	-0.079	&	0.014	&		&		&		&		\\
319.51548	&	-15.97303	&	53226.540177	&	2.39	&	6.87	&	17.151	&	20.531	&	0.005	&	19.769	&	0.002	&	19.490	&	0.010	&	19.097	&	0.002	&	-0.083	&	0.004	&		&		&		&		\\
319.69070	&	-15.97569	&	53226.540177	&	3.02	&	9.71	&	14.657	&	19.295	&	0.004	&	18.576	&	0.005	&	18.222	&	0.006	&	17.868	&	0.002	&	-0.061	&	0.005	&		&		&		&		\\
319.45615	&	-16.33313	&	53226.540177	&	2.99	&	12.99	&	18.454	&	23.144	&	0.004	&	22.319	&	0.009	&	21.933	&	0.025	&	21.498	&	0.031	&	0.036	&	0.015	&		&		&		&		\\
319.69838	&	-16.32601	&	53226.540177	&	2.51	&	0.48	&	20.177	&	24.522	&	0.018	&	23.069	&	0.406	&	23.014	&	0.050	&	22.693	&	0.037	&	0.247	&	0.303	&		&		&		&		\\
319.69109	&	-16.28867	&	53226.540177	&	2.63	&	6.04	&	20.354	&	23.898	&	0.198	&	23.519	&	0.050	&	23.163	&	0.021	&	22.787	&	0.005	&	-0.300	&	0.147	&		&		&		&		\\
319.85657	&	-16.27948	&	53226.540177	&	3.12	&	0.50	&	19.148	&	23.844	&	0.026	&	23.243	&	0.003	&	22.847	&	0.027	&	22.483	&	0.043	&	-0.115	&	0.024	&		&		&		&		\\
319.85490	&	-16.26903	&	53226.540177	&	2.68	&	2.55	&	20.756	&	24.994	&	0.056	&	24.023	&	0.117	&	23.626	&	0.089	&	23.131	&	0.099	&	0.147	&	0.167	&		&		&		&		\\
319.56316	&	-16.23162	&	53226.540177	&	3.22	&	4.24	&	15.270	&	20.221	&	0.006	&	19.547	&	0.020	&	19.088	&	0.012	&	18.382	&	0.013	&	-0.019	&	0.021	&		&		&		&		\\
319.42440	&	-16.20991	&	53226.540177	&	2.96	&	0.96	&	20.773	&	25.443	&	0.034	&	24.592	&	0.056	&	24.029	&	0.040	&	23.412	&	0.061	&	0.179	&	0.079	&		&		&		&		\\
320.79963	&	-16.11936	&	53227.544813	&	2.99	&	1.87	&	18.126	&	22.710	&	0.008	&	21.997	&	0.018	&	21.638	&	0.034	&	21.361	&	0.009	&	-0.062	&	0.041	&		&		&		&		\\
320.56040	&	-16.08520	&	53227.544813	&	2.04	&	5.53	&	21.185	&	23.541	&	0.010	&	22.824	&	0.023	&	22.194	&	0.017	&	21.742	&	0.022	&	0.132	&	0.024	&		&		&		&		\\
320.64800	&	-16.03000	&	53227.544813	&	2.87	&	17.42	&	17.010	&	21.444	&	0.009	&	20.651	&	0.006	&	20.265	&	0.005	&	19.901	&	0.006	&	0.014	&	0.009	&		&		&		&		\\
320.27458	&	-16.02996	&	53227.544813	&	3.05	&	7.99	&	17.404	&	22.289	&	0.046	&	21.376	&	0.014	&	20.940	&	0.011	&	20.564	&	0.024	&	0.134	&	0.035	&		&		&		&		\\
320.40652	&	-16.01986	&	53227.544813	&	2.29	&	12.74	&	20.260	&	23.424	&	0.047	&	22.600	&	0.020	&	22.209	&	0.028	&	21.855	&	0.002	&	0.039	&	0.047	&		&		&		&		\\
320.31635	&	-16.00107	&	53227.544813	&	2.25	&	0.53	&	19.489	&	22.591	&	0.019	&	21.726	&	0.000	&	21.300	&	0.008	&	20.938	&	0.004	&	0.093	&	0.026	&		&		&		&		\\
320.82385	&	-15.98608	&	53227.544813	&	2.44	&	3.24	&	18.920	&	22.623	&	0.021	&	21.652	&	0.010	&	21.032	&	0.009	&	20.621	&	0.007	&	0.305	&	0.021	&		&		&		&		\\
320.71048	&	-15.93982	&	53227.544813	&	2.64	&	4.53	&	19.094	&	23.258	&	0.031	&	22.283	&	0.004	&	21.818	&	0.010	&	21.453	&	0.019	&	0.198	&	0.028	&		&		&		&		\\
320.65166	&	-15.91541	&	53227.544813	&	2.33	&	7.10	&	20.093	&	23.306	&	0.004	&	22.550	&	0.007	&	22.065	&	0.029	&	21.109	&	0.511	&	0.057	&	0.006	&		&		&		&		\\
320.62216	&	-15.91270	&	53227.544813	&	3.13	&	3.21	&	16.622	&	21.451	&	0.007	&	20.738	&	0.007	&	20.321	&	0.004	&	19.957	&	0.001	&	-0.021	&	0.008	&		&		&		&		\\
320.39658	&	-15.87950	&	53227.544813	&	2.56	&	2.40	&	19.464	&	23.241	&	0.008	&	22.462	&	0.062	&	22.075	&	0.010	&	21.649	&	0.004	&	0.004	&	0.062	&		&		&		&		\\
320.41451	&	-15.86789	&	53227.544813	&	2.72	&	3.93	&	18.944	&	22.866	&	0.126	&	22.304	&	0.055	&	21.900	&	0.047	&	21.499	&	0.038	&	-0.137	&	0.119	&		&		&		&		\\
320.39999	&	-15.85829	&	53227.544813	&	3.10	&	0.75	&	18.993	&	23.920	&	0.041	&	23.057	&	0.044	&	22.648	&	0.020	&	22.590	&	0.149	&	0.079	&	0.048	&		&		&		&		\\
320.70450	&	-15.85786	&	53227.544813	&	2.64	&	1.98	&	19.097	&	23.084	&	0.035	&	22.283	&	0.010	&	21.895	&	0.013	&	21.556	&	0.000	&	0.021	&	0.040	&		&		&		&		\\
320.66966	&	-15.78510	&	53227.544813	&	3.22	&	1.12	&	15.826	&	20.814	&	0.018	&	20.095	&	0.006	&	19.707	&	0.004	&	19.393	&	0.004	&	-0.037	&	0.014	&		&		&		&		\\
320.65956	&	-15.74312	&	53227.544813	&	2.91	&	24.42	&	17.230	&	21.705	&	0.018	&	20.951	&	0.000	&	20.661	&	0.006	&	20.222	&	0.001	&	-0.082	&	0.015	&		&		&		&		\\
320.51081	&	-15.74918	&	53227.544813	&	2.69	&	6.83	&	19.273	&	23.438	&	0.014	&	22.559	&	0.000	&	22.276	&	0.032	&	21.796	&	0.000	&	0.002	&	0.018	&		&		&		&		\\
320.65568	&	-15.73321	&	53227.544813	&	2.51	&	3.04	&	19.194	&	23.230	&	0.011	&	22.099	&	0.010	&	21.434	&	0.127	&	21.270	&	0.001	&	0.450	&	0.027	&		&		&		&		\\
320.32995	&	-15.71095	&	53227.544813	&	3.08	&	12.77	&	18.356	&	23.137	&	0.015	&	22.386	&	0.037	&	22.007	&	0.007	&	21.630	&	0.004	&	-0.021	&	0.032	&		&		&		&		\\
320.30033	&	-15.68513	&	53227.544813	&	3.15	&	10.08	&	18.531	&	23.528	&	0.019	&	22.691	&	0.000	&	22.247	&	0.010	&	21.592	&	0.191	&	0.086	&	0.030	&		&		&		&		\\
\enddata
\tablecomments{$^a$ Estimated by using Equation 6, 7,8 and 9 under assuming e=0.}
\tablecomments{$^b$ See Equation 13.}
\tablecomments{$^c$ Orbital elements of known asteroids are provided by JPL Small-Body Database.}

\end{deluxetable}

\subsection{Flux Calibration}
The Suprime-Cam image data were calibrated chip by chip in every exposure by identifying Pan-STARRS 1 (hereafter PS1) catalogue stars in the observed field;
the data were calibrated using ``uber-calibration" \citep{mag13}. 
   Uber-calibration is an algorithm used to photometrically calibrate
   wide-field optical imaging surveys, and it was first applied to the Sloan
   Digital Sky Survey imaging data. It can be used to simultaneously solve for
   the calibration parameters and relative stellar fluxes through the use of 
   overlapping observations \citep{pad08}.  
Those uber-calibrated catalogues have a relative precision (compared with
the SDSS) of $<$ 10 mmag in $g_{P1}$, $r_{P1}$, and $i_{P1}$, and approximately 
10 mmag in $z_{P1}$ and $y_{P1}$  \citep{sch12}. 
Since we used the $B$, $V$, $R$, $I$ filters in our survey, we transferred those PS1 catalogue stars from the PS1 $g_{P1}r_{P1}i_{P1}$ photometry system to our Johnson-Cousins $BVRI$ system by using the transformation equation and parameters obtained by \citet{ton12}.
The transformation error was approximately 0.03 in the $B$-band and 0.01 to 0.02 in the $V$-, $R$-, $I$-bands.

\subsection{Trail Fitting}

Two problems occur when using the traditional photometry methods for asteroid flux measurements: 
First, the flux would be contaminated by nearby stars if aperture photometry is used to measure the asteroid flux within crowded fields. 
Second, PSF fitting fails to yield accurate results if the asteroid image is not point-source-like object under long exposure. 
Therefore, we applied the trail fitting technique to measure the asteroid flux.

The trail function chosen is an axisymmetric Gaussian PSF-convolution trail function given in equation (3) of \citet{ver12}:
\begin{equation}
g[x^\prime,y^\prime] = b +
  \frac{F}{L}\frac{1}{2\sigma\left(2\pi\right)^{1/2}}
  \exp{
    \left[-\frac{{y^\prime}^2}{2\sigma^2}\right]
  }
    \left\{
      erf\left[\frac{x^\prime+\frac{1}{2}L}{\sigma\sqrt{2}}\right] - 
      erf\left[\frac{x^\prime-\frac{1}{2}L}{\sigma\sqrt{2}}\right]
      \right\}, \label{eq:erf}
\end{equation}
where
\begin{equation}
erf[z] = \frac{2}{\pi}\int_0^z \exp{\left[ t^2 \right]} dt
\end{equation}
and
\begin{eqnarray}
  x^\prime &=&  (x-x_c)\cos{\theta} + (y-y_c)\sin{\theta},\\
  y^\prime &=& -(x-x_c)\sin{\theta} + (y-y_c)\cos{\theta}, 
\end{eqnarray}
where $b$ is the background level, $L$ is the length of the trail, $F$ is the total integrated
flux in the trail, $\sigma$ is the standard deviation of the PSF Gaussian, $x_c$ and $y_c$ are the coordinates of the centroid of the trial.
$\theta$ is the angle between the long axis of the trail and the $x$~axis.

\begin{figure}
%\epsscale{0.3}
\includegraphics[width = .5\textwidth]{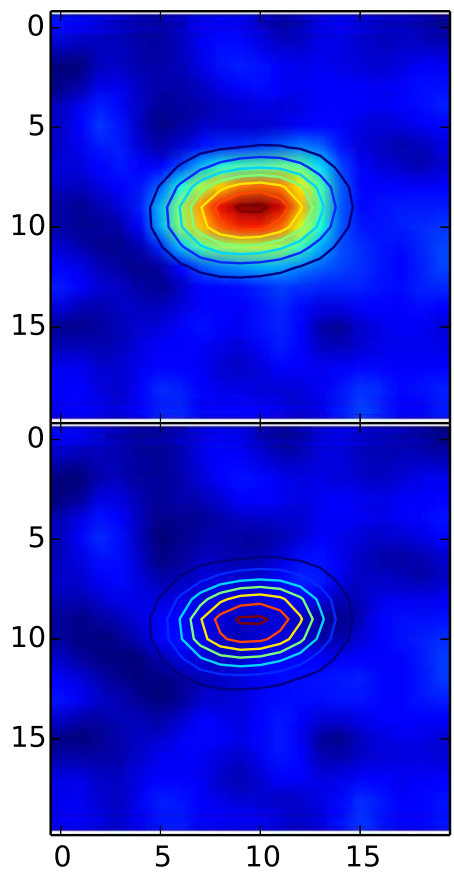}
\caption{An example of an asteroid trail in our sample before (top) and after (bottom) subtracting its fitting result (contour). \label{fig1}}
\end{figure}

We used the Levenberg-Marquardt least-squares fitting technique to minimize the variance between the image and trail function. Figure~\ref{fig1} shows an example of an asteroid trail and its fitting residue. Clearly, the asteroid trail can be completely subtracted by using the trail function.
The photometric error of the asteroid flux was estimated from the flux difference between two consecutive exposures of the asteroid; we considered the flux difference to be closer to the actual uncertainty compared with the estimation derived from the fitting result. We expect that the SNR of trail fitting photometry decreases with ($2.35\sigma$/($2.35\sigma+L$))$^{0.5}$. Thus the faster mover have larger photometry uncertainty.  Fortunately, there is no systematic effect related to the trail length (See \citet{ver12} for more detail), and the photometry results should be independent with the semi-major axis of asteroids.

\subsection{Detection Efficiency and De-biasing}

For testing the detection efficiency of moving objects in the observations, we planted 500 non-trailing artificial stars with a brightness between 22.5 to 25.5 magnitude in each CCD chip. The artificial objects were generated by $IRAF$ task $DAOPHOT$ by modeling the stars in each CCD chip in each exposure.
The limiting magnitude also depends on the trailing rate of asteroids. Assuming that the moving rate of asteroids is about 0.01''/s in average.  We expect that the trail length is around 1.2'' in the 120s exposure. For the average 0.7'' seeing size of our data, the SNR of the asteroid trail is roughly 60$\%$ of the non-trailing source with the same brightness. Therefore, we increased the detection threshold of artificial non-trailing stars to 5 sigma to simulate the limiting magnitude of 3 sigma asteroid detections.

We counted the number of detected artificial objects with a function of magnitude and plotted a diagram of magnitude vs. fractional detection as shown in Figure ~\ref{fig2}, and fitted a detection efficiency function, which was a function with double hyperbolic tangents \citep{pet06}, to the test result:

\begin{figure}
%\epsscale{0.5}
\includegraphics[width = .7\textwidth]{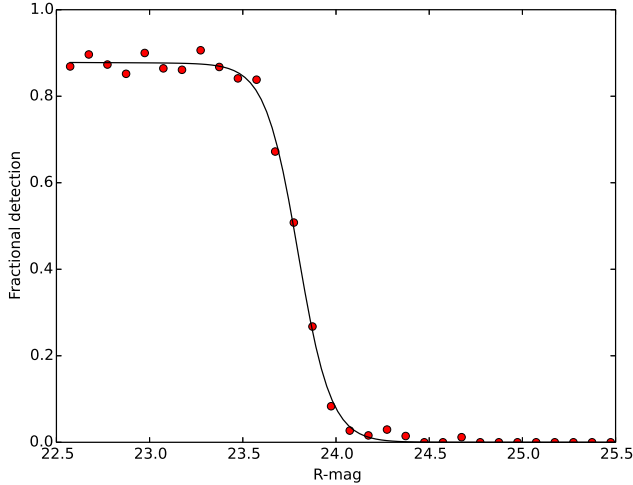}
\caption{An example of fractional detection as a function of the R-magnitude. The best-fit detection efficiency function is illustrated. \label{fig2}}
\end{figure}

\begin{equation}
\eta(R) = \frac{A}{4}[1-tanh(\frac{R-R_c}{\Delta_1})][1-tanh(\frac{R-R_c}{\Delta_2})]
\end{equation}
Here, the fitted parameters A, R$_c$, $\Delta_1$ and $\Delta_2$ are the filling factor (or maximal efficiency), roll-over magnitude ($50\%$ of the maximal efficiency), and widths of the two components, respectively.
An exemplary result for the R-band, Chip2, took on  August 10, 2004 was the $50\%$ detectability of approximately 23.7 magnitude, and the filling factor was approximately 83$\%$ (see Figure~\ref{fig2}).

Using this detection efficiency function, we assigned a weight for each asteroid detection; the weight was the multiplicative inverse of the product of the detection efficiency of every chip and exposure of the asteroid that passed. The weight of the object with the highest detectability is 1, and fainter objects, which have a low detection probability, have a higher weighting value. Therefore, we eliminated the observational bias in the detection of objects and tested the completeness of the survey.

\section{Results}

\subsection{Absolute Magnitude and Completeness}
We extracted a total of 150 asteroids with 12 detections from the six fields and measured their apparent velocities. Assuming that their orbital eccentricities were zero and they located around opposition during the observations, we estimated the semi-major axes and inclinations by using the following equations \citep{bow79, nak02, yos07}:

\begin{eqnarray}
a &=& \frac{1}{2\gamma}(\gamma-2k\lambda\pm\sqrt{\mid\kappa\mid}), \\
\tan{I} &=& \frac{\mid\beta\mid}{\lambda+k/(a-1)}, \\
\gamma &=& \lambda^2 + \beta^2, \\
\kappa &=& \gamma^2 - 4k\lambda\gamma - 4k^2\beta^2,
\end{eqnarray}
where $\lambda$ and $\beta$ are the moving rate  along the ecliptic longitude and latitude, respectively;  $a$ and $I$ are the semi-major axis and inclinations, respectively; and  k is the Gaussian gravitational constant.
The semi-major axis and inclinations estimated using the aforementioned equations included errors of approximately 0.1 AU and 5$^\circ$ because the actual eccentricities of the asteroids were not zero. We compared the orbital elements of known asteroids in Table~\ref{tab2} with our estimated elements. The result shows that the estimate elements are generally accurate; the difference between known and estimated semi-major axes, inclinations are less than 0.1 AU, 2$^\circ$, respectively.

The absolute magnitude $(H_V)$ of each asteroid at opposition can be estimated using the following equation:
\begin{equation}
H_V = m - 5  log(\Delta r) 
\end{equation}
where m is the apparent magnitude of the asteroid and $\Delta$ and r are geocentric distance and heliocentric distance, respectively. For simplicity,  $r = a$ and $\Delta = a - 1$ because of the $e = 0$ and nearly opposition assumptions.

\begin{figure}
%\epsscale{1}
\includegraphics[width = .7\textwidth]{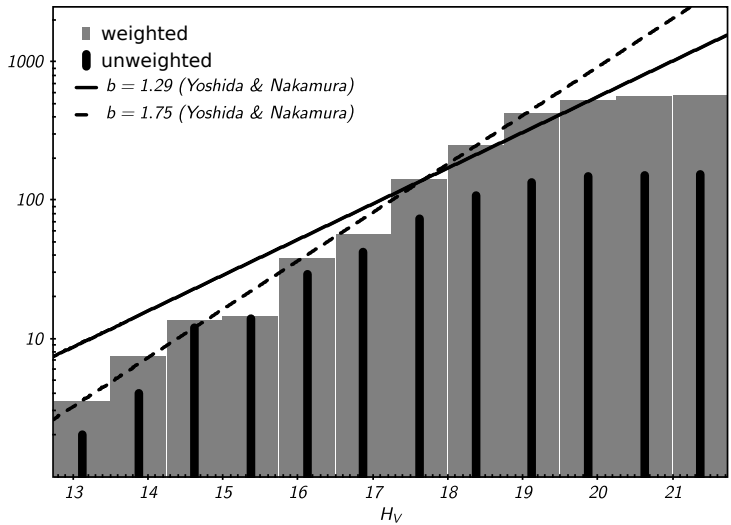}
\caption{Weighted (Grey) and unweighted (black) cumulative absolute magnitude ($H_V$) distributions of the asteroids sample. Solid and dashed lines show the power laws distribution with power law indices of 1.29 (for the $H_V$ range 17.8 to 20.2 mag) and 1.75 (14.6 to 17.4 mag), respectively. \label{fig3}}
\end{figure}

Figure~\ref{fig3} shows the absolute magnitude ($H_V$) distribution of our samples. The black spikes are the raw $H_V$ distribution, and the gray boxes shows the weighted result.
Based on the results of  \citet{yos07}, we plotted two power laws (N ($< H_V$) $\propto 10^{\alpha H_V} $) to the $H_V$ distribution, with power law indices b of 1.29 (for  the $H_V$ range from 17.8 to 20.2 mag) and 1.75 (14.6 to 17.4 mag). Here b = 5 $\alpha$.
This plot clearly indicates that our samples have power law indices consist with the results of  \citet{yos07} and satisfactory completeness of up to $H_V = 20$, which corresponds to asteroids with diameters smaller than 590 meter for the C-complex asteroids (assuming an albedo of 0.05) and smaller than 270 meter for the S-complex asteroids (assuming an albedo of 0.25).

\subsection{Colors and Estimated Taxonomic Types}

\begin{figure}
%\epsscale{0.5}
\includegraphics[angle=0, width = .7\textwidth]{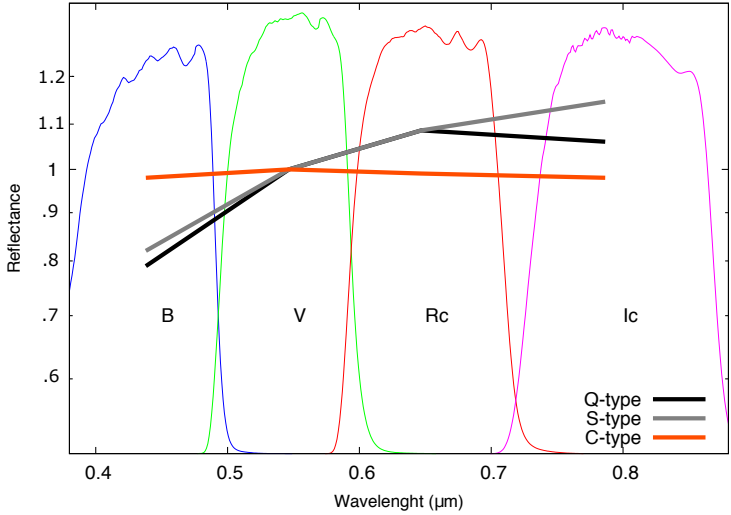}
\caption{Spectrophotometry of three types of asteroid spectra: S-complex, C-complex and V/Q/R-type-like asteroids. The background shows the transmission curves of the Suprime-Cam filter system. \label{fig4}}
\end{figure}

From our sample (150 asteroids), 75 asteroids with the $H_V$ range from 16 to 20 magnitude, the color errors smaller than 0.1, 0.2 $< V-R < 0.6$ and 0.6 $< B-V <$ 1 was selected. These asteroids have satisfactory $H_V$ completeness and photometric accuracy which made them suitable for taxonomic estimations. The color errors are propagated from the uncertainly of BVRI photometry of asteroids in Table~\ref{tab2}.
Figure~\ref{fig4} shows the relative reflectances in our asteroid samples overlap with the transmission curve of the Suprime-Cam filter system. The C-complex asteroids generally have a flat reflectance across the four filters, and the S-complex and V/Q/R-type asteroids exhibit similar slopes in the $B, V, R$ region but different reflectances near the I-band; the V/Q/R-type asteroids demonstrated a stronger absorption feature at $\sim$ 1 $\mu$m. Therefore, while $B-V$ and $V-R$ color are used to separate S-complex and C-complex asteroids, the $R-I$ color is needed to distinguish S-type asteroids from V/Q/R-type-like asteroids in the following sense. The S-type asteroids are those with $R - I > (R - I)_{\odot}$ (0.332 $\pm$ 0.008, according to \citet{hol06}); otherwise the V/Q/R-type-like asteroids.

\begin{figure}
%\epsscale{0.5}
\includegraphics[angle=270, width = .7\textwidth]{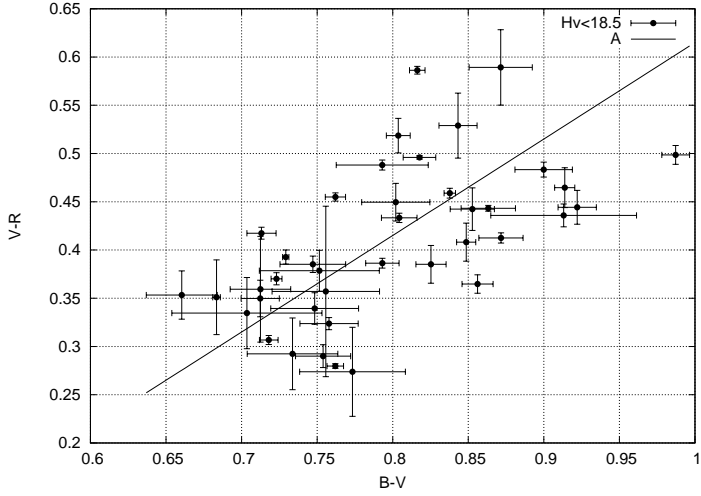}
\caption{The $B-V$ and $V-R$ color-color diagram for our large asteroid sample ($H_V < 18.5$). The solid line shows the direction of new axis (A) of the color-color distribution. \label{fig5}}
\end{figure}

We also separated our sample into large and small asteroids by $H_V = 18.5$, which approximately corresponds to a diameter of 1 km for C-complex asteroids.
Figure~\ref{fig5} shows $B-V$ vs. $V-R$ color-color diagram for part of our large asteroid sample ($H_v < 18.5$). The bimodular distribution corresponds to the color difference between the C-complex and S-complex asteroids. 

Moreover, we identified principal components (the uncorrelated variables) for the large asteroid sample in $B-V$ and $V-R$ space by prcincipal component analysis to separate S-complex asteroids and C-complex asteroids and resulted two linear combinations of $B-V$ and $V-R$  colors:
\begin{equation}
PC1 \equiv 0.7071(B-V) + 0.7071(V-R),
\end{equation}
and
\begin{equation}
PC2 \equiv 0.7071(B-V) - 0.7071(V-R)
\end{equation}
The result shows that $B-V$ and $V-R$ axes become PC1 and PC2, respectively, after rotating 45$^\circ$ counterclockwise.

Furthermore, there is a dip at PC1 = 0.82 in the histogram of PC1 and can be the boundary of C and S-complex asteroids. Therefore, we define a new axis `$A$':
\begin{equation}
A \equiv 0.7071(B-V) + 0.7071(V-R) - 0.82
\end{equation}
Asteroids with $A >$ 0 should belong to S-complex, otherwise are C-complex.
The new axis 'A' is shown as a solid line in Figure~\ref{fig5}.

\begin{figure}
%\epsscale{0.5}
\includegraphics[angle=270, width = .7\textwidth]{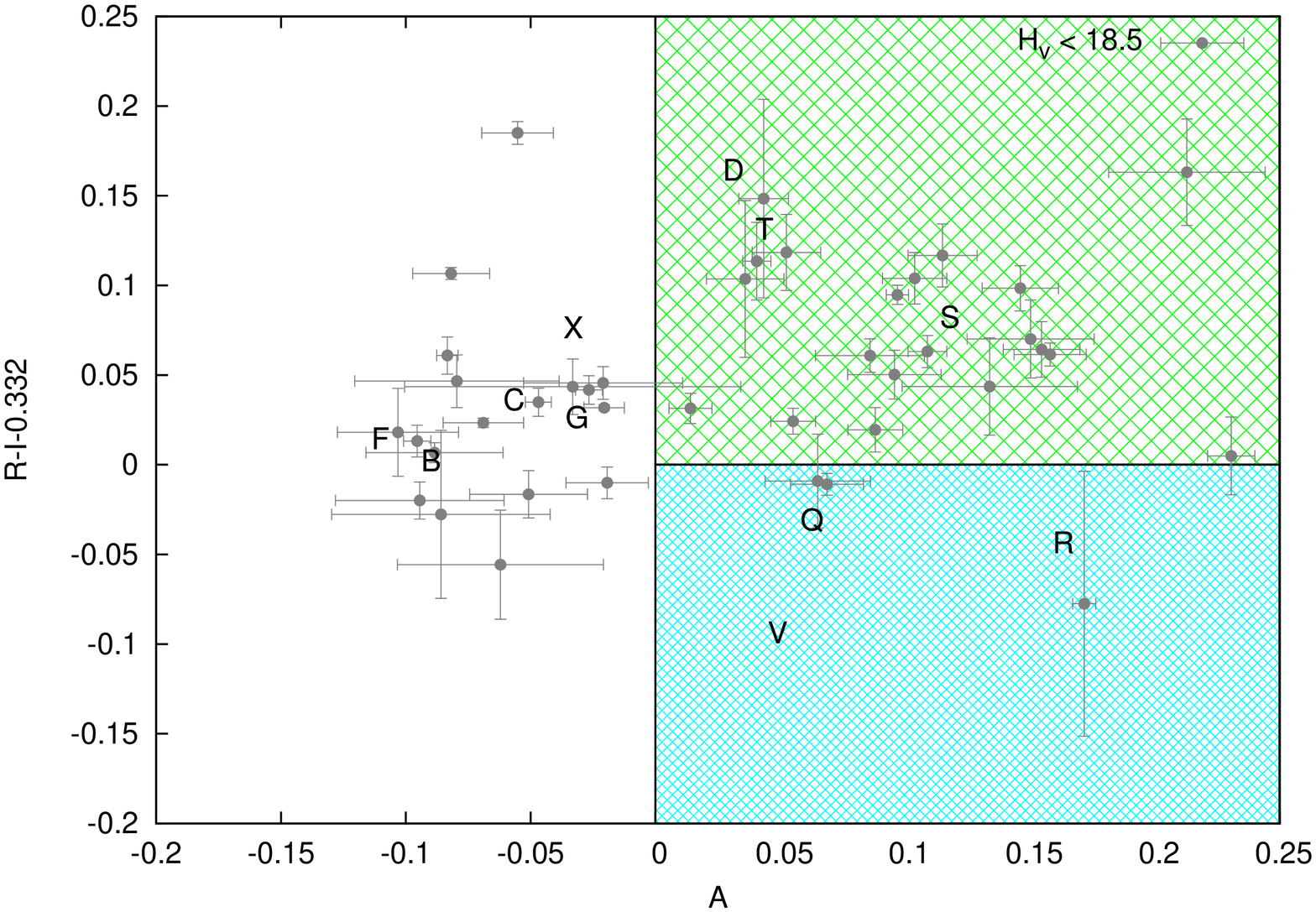}
\includegraphics[angle=270, width = .7\textwidth]{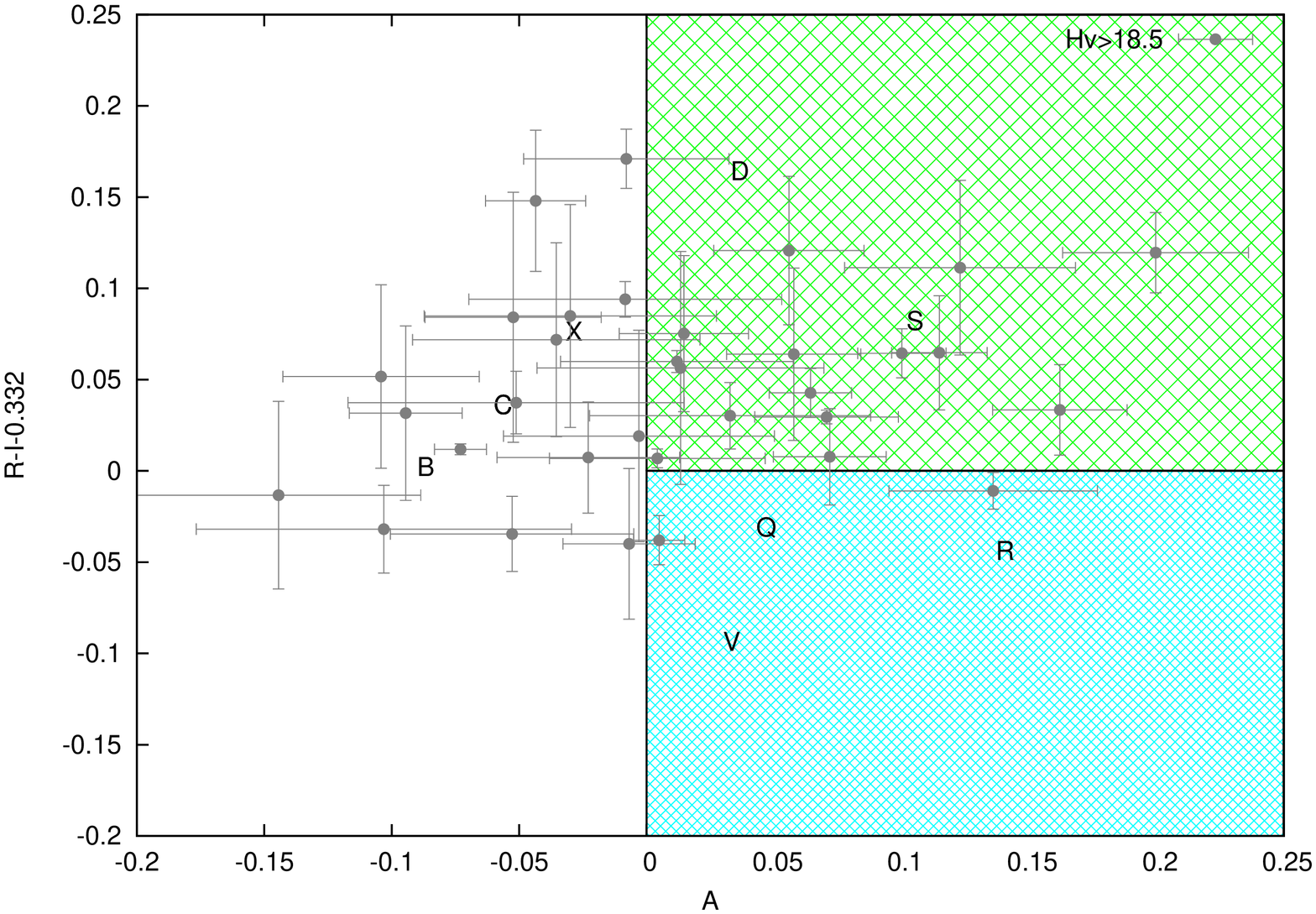}
\caption{The $R - I$ vs. $A$ color-color diagrams of the large (upper) and small (bottom) asteroid samples in the present Subaru data. The regions corresponding to three types of asteroids are shown in the plots. The average colors of B, C, D, F, G, S, T, V, Q, R and X-type asteroids from \citet{dan03} are also marked. \label{fig6}}
\end{figure}

Figure~\ref{fig6} shows the color-color diagram of axis $A$ vs. $R-I$.  Each dot represents an asteroid. A value of 0.332, which is corresponding to the solar color, was subtracted from the $R-I$ axis to separate S-complex and V/Q/R-type.
Two major points can be made based on Figure~\ref{fig6}: (1) only two objects can be certified as Q-type MBA candidates;  (2) the small sample ($H_V > 18.5 $) appeared to be more concentrated at $A \approx 0$, though the C- and S-complexes are clearly separated in large sample ($Hv < 18.5$).

\subsection{Fraction of Each Estimated Taxonomic Type}

\begin{figure}
%\epsscale{0.5}
\includegraphics[angle=270, width = .7\textwidth]{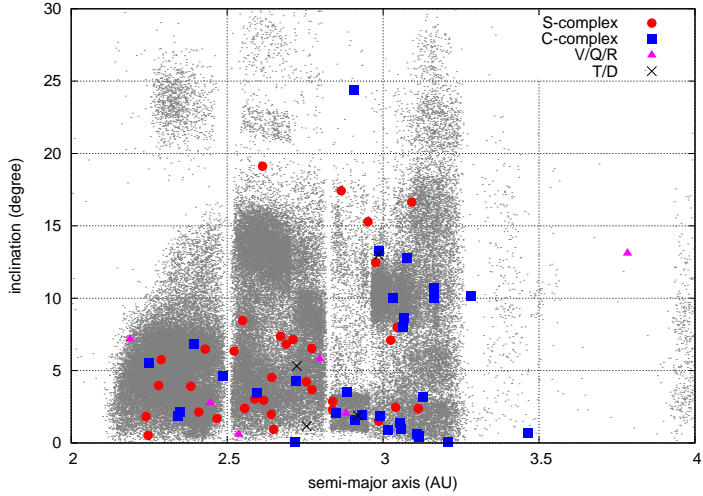}
\caption{Semi-major axis (a) vs inclination (i) distribution of the 75 asteroids with estimated taxonomic type. Gray dots show the a, i distribution of of known asteroids. \label{fig7}}
\end{figure}

To estimate the taxonomic type of asteroids, we considered $A < 0$ as C-complex asteroids, $A > 0$ and $R - I -0.332 > 0$ as S-complex and  $A > 0$ and $R - I - 0.332 < 0$ as V/Q/R like asteroids. Four objects locate around D/T type region were removed by visually selection. Figure~\ref{fig7} shows the semi-major axis vs inclination distribution of the 75 asteroids with estimated taxonomic type. The distribution looks normal; there are more C-complex than S-complex asteroids in outer main-belt. There is a possible V/Q/R-type candidate located on 3.8 AU. It could because of wrong estimation in semi-major axis.

\begin{figure}
%\epsscale{0.5}
\includegraphics[angle=270, width = .7\textwidth]{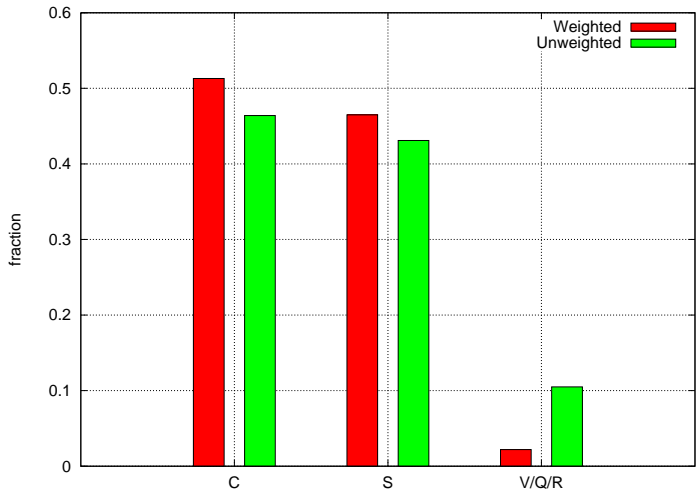}
\caption{Histograms of the fractions of estimated asteroid spectral types. C and S are C-complex and S-complex asteroids, respectively, and V/Q/R indicates V/Q/R-type-like asteroids. \label{fig8}}
\end{figure}

To calculate the fraction of each estimated taxonomic type, we took the uncertainly of boundaries of $A$ and $R - I$ axes, which are 0.02 from the bin-size of $A$ histogram and 0.008 from the uncertainty of $(R - I)_{\odot}$, respectively, and the uncertainly of asteroid color measurements into account. We use 2D gaussian distribution with a covariance matrix equal to 
\[\left( \begin{array}{cc}
$A$_{err}^2 & 0 \\
0 & $(R - I)$_{err}^2 
\end{array}\right)\]
to represent the probability distribution of each object in $A$, $R - I$ space. Therefore, the fraction of a object in a specific taxonomic type is its probability distribution multiply a Complementary Error Function, which is the cumulative function of a gaussian distribution with the mean value equal to boundary value and standard deviation equal to the uncertainty of boundary. Figure~\ref{fig8} shows the fraction of asteroid types in our weighted and unweighted sample.

Since our sample exhibits a $H_V$ range similar to that of the near-Earth asteroid sample used by \citet{dan03}, it is worth to compare the fraction of asteroid's type to that of NEA population. The Q/S ratio in our observation is less than 0.05 in the main-belt region. By contrast, the Q/S ratio in the near earth space is about 0.5 \citep{bin04} to 2 \citep{dan03}, which is much higher than in the main-belt.

\begin{figure}
%\epsscale{0.5}
\includegraphics[angle=270, width = .7\textwidth]{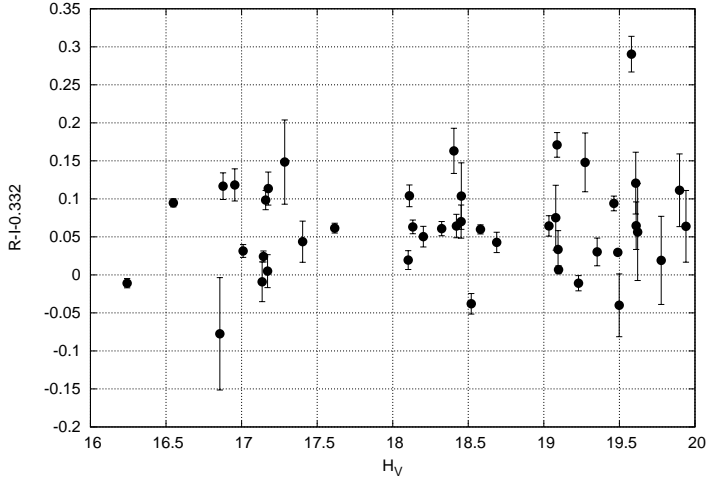}
\caption{Variation of $R - I$ color (an indicator of absorption band depth) as a function of the absolute magnitude, $Hv$, for the observed S-complex MBAs.
\label{fig9}}
\end{figure}

Finally, we tested the correlation between $H_V$ and $R - I$ of S-complex asteroids. As shown in Figure~\ref{fig9}, unlike the result of \citet{dan03} in which the absorption band depth was correlated with S-complex NEA size, we did not find any evidence that $R - I$ color has significant correlation with the size of S-complex MBA. 

\section{Discussion}

There are only two possible Q-type asteroids candidates in our km to sub-km sized MBA sample (see Figure~\ref{fig6}). This fact indicates that the Q-type asteroids are rare in the main-belt. The Q/S ratio is less than 0.05, which is significantly lower than the value of NEAs ($\sim 0.5$ in \citet{bin04} and $\sim 2$ in \citet{dan03}). This result is also comparable with the Q-type-like fraction in \citet{car10}, which the main belt Q/S ratio is $325/3784 \sim 0.09$ for the good classified asteroids in SDSSMOC4.

Since most of S-complex MBAs are weathered, we can estimate the upper limit of space weathering timescale in main belt by using the collisional size-age relation in \citet{bot05a} and \citet{wil11}:
\begin{equation}
T_s(yr)= 3.23 \times 10^{(11.0 - 0.18H_v)}
\end{equation}
The H$_v$ range of our MBA samples are between 16 to 20 magnitude, and their corresponding collisional ages are between $8 \times 10^7$ to $4 \times10^8$ years. We can safely conclude that the space weathering timescale in main belt should be less than $10^8$ years.

The lack of Q-type MBAs also suggests two facts: 1. Most of the Q-type NEAs did not come from main-belt, they must form in-situ, and 2. collision is not the main mechanism of the formation of Q-type NEAs due to the collision rate being lower in the near Earth region.
There must exist other mechanisms to generate such large amount of Q-type NEAs, and these mechanisms are more effective in the near Earth region than the main-belt. The planetary encounter model \citep{bin10, nes10} advocating the recent resetting of  S-type asteroid surfaces by the effects of tidal stress is one of the possible mechanisms. 

Another possible mechanism that could be responsible for the formation of Q-type NEAs is YORP effect spin-up induced rotational-fission or surface re-arrangement of asteroids. The acceleration rate of asteroid spin by the YORP effect is inversely proportional to the square of semi-major axis and more effective in the near Earth region than in the main-belt due to the smaller heliocentric distance \citep{rub00, sch07}. Thus, if rotational-fission mechanism or rotational re-arrangement is also several times more effective to create Q-type NEAs than Q-type MBAs, it may be able to explain why the Q/S ratio in NEAs is about 10 to 40 times larger than Q/S ratio in MBAs.

The YORP spin-up can also explain the existence of main-belt Q-type asteroids (see \citet{pol14} for detail).  It indicates the size-color (or size-S/Q ratio) dependence of S-complex MBAs, because the smaller asteroid is easier to be accelerated to near the break-up limit resulting in the Q-type-like color. However, such relation is not shown in our sample. There are two possible effects that may cause the lack of size dependence:\\
1. The ``secondary fission'' of rotational-fission models provided by \citet{jac11} may be the more likely model to create main-belt Q-type asteroids. This model implies the destruction of the secondary of pair asteroids by primary's tidal forces. For the km to sub-km sized asteroid pair, the gravity may be too small to deform secondary and produce Q-type surface. \\
2. A size-dependent strength for asteroids in addition to gravity \citep{hol07} can prevent the break-up of small asteroids. The existence of sub-km sized super-fast rotator, such as (29075) 1950 DA \citep{roz14} and (335433) 2005 UW163 \citep{cha14}, might be the evidence of the existence of this internal strength.

The other mechanism is thermal degradation of the rocks on asteroid surface from \citet{del14}. This process is strongly dependent on the value of diurnal temperature difference, which is a function of perihelion distance; it takes $\sim 10^5$ yrs in near Earth region and $\sim 10^7$ yrs in the main-belt to break $90\%$ of 3 centimeter diameter size rocks. If thermal degradation dominates the formation of Q-type asteroids, space weathering must have timescale $< 10^7$yrs to keep low Q/S ratio in the main-belt.

%If rotational-fission mechanism can not present the size-color dependence of asteroid, the size-color dependence/independence in S-complex NEAs/MBAs would indicate that at least two different mechanisms should respond to the formation of Q-type MBAs and NEAs; the tidal effect during planetary encounters is favorable for the Q-type NEAs formation and the rotational-fission is more suitable for the Q-type MBAs formation.
%The further investigations of size-color dependence of MBAs and NEAs will give the better answer of the question.

%We found that the smaller C-complex asteroids appear redder than the larger ones (Figure~\ref{fig9}). However, the correlation disappear once we test their age-spectral slope relation. At least in the observational point of view, the space-weathering effect on C-complex asteroids remains unclear.

\section{Summary}
We surveyed kilometer- to sub-kilometer-sized asteroids in the main belt by using the Subaru telescope. A total of 150 asteroids with BVRI colors were detected and 75 of them exhibited satisfactory photometry accuracy. The main results can be summarized as follows:\\
1. Q-type asteroids are rare in the main-belt; only two Q-type candidates were detected in our sample, and the Q-type to S-type ratio is less than 0.05 in main-belt.\\
2. Unlike the size-color dependence of NEAs found by \citet{dan03}, we did not found any evidence of that in MBA population. \\
3. The space weathering timescale in the main belt should be less than $10^8$ years. \\
4. Re-arrangement of surface material of S-type asteroid by tidal stress during planetary encounters and thermal degradation are possible mechanisms of Q-type NEAs formation. YORP spin-up induced rotational-fission or surface re-arrangement of asteroids could be responsible for both Q-type MBAs and NEAs formation.\\
\\
Acknowledgments
\\

We also acknowledge the anonymous referees' useful suggestions for improving the manuscript.
This work is based on data collected at Subaru Telescope, which is operated by the National Astronomical Observatory of Japan. This work was supported in part by NSC Grant: NSC 101-2119-M-008-007-MY3 and NSC 102-2119-M-008-001. 
The Pan-STARRS1 Surveys (PS1) have been 
made possible through contributions by the Institute for Astronomy, the 
University of Hawaii, the Pan-STARRS Project Office, the Max-Planck 
Society and its participating institutes, the Max Planck Institute for 
Astronomy, Heidelberg and the Max Planck Institute for Extraterrestrial 
Physics, Garching, The Johns Hopkins University, Durham University, 
the University of Edinburgh, the Queen's University Belfast, the 
Harvard-Smithsonian Center for Astrophysics, the Las 
Cumbres Observatory Global Telescope Network Incorporated, the 
National Central University of Taiwan, the Space Telescope Science Institute, and the National 
Aeronautics and Space Administration under Grant No. NNX08AR22G issued 
through the Planetary Science Division of the NASA Science Mission 
Directorate, the National Science Foundation Grant No. AST-1238877,
the University of Maryland, Eotvos Lorand University (ELTE),
and the Los Alamos National Laboratory.


\begin{thebibliography}{00}

\bibitem[Bernstein and Khushalani(2000)]{ber00} Bernstein, 
G., Khushalani, B.\ 2000.\ Orbit Fitting and Uncertainties for Kuiper Belt 
Objects.\ The Astronomical Journal 120, 3323-3332. 

\bibitem[Binzel et al.(1993)]{bin93} Binzel, R.~P., Xu, S., 
Bus, S.~J., Skrutskie, M.~F., Meyer, M.~R., Knezek, P., Barker, E.~S.\ 
1993.\ Discovery of a Main-Belt Asteroid Resembling Ordinary Chondrite 
Meteorites.\ Science 262, 1541-1543. 


\bibitem[Binzel et al.(1996)]{bin96} Binzel, R.~P., Bus, 
S.~J., Burbine, T.~H., Sunshine, J.~M.\ 1996.\ Spectral Properties of 
Near-Earth Asteroids: Evidence for Sources of Ordinary Chondrite 
Meteorites.\ Science 273, 946-948. 


\bibitem[Binzel et al.(2001)]{bin01} Binzel, R.~P., Harris, 
A.~W., Bus, S.~J., Burbine, T.~H.\ 2001.\ Spectral Properties of Near-Earth 
Objects: Palomar and IRTF Results for 48 Objects Including Spacecraft 
Targets (9969) Braille and (10302) 1989 ML.\ Icarus 151, 139-149. 

\bibitem[Binzel et al.(2004)]{bin04} Binzel, R.~P., Rivkin, 
A.~S., Stuart, J.~S., Harris, A.~W., Bus, S.~J., Burbine, T.~H.\ 2004.\ 
Observed spectral properties of near-Earth objects: results for population 
distribution, source regions, and space weathering processes.\ Icarus 170, 
259-294.

\bibitem[Binzel et al.(2010)]{bin10} Binzel, R.~P., 
Morbidelli, A., Merouane, S., DeMeo, F.~E., Birlan, M., Vernazza, P., 
Thomas, C.~A., Rivkin, A.~S., Bus, S.~J., Tokunaga, A.~T.\ 2010.\ Earth 
encounters as the origin of fresh surfaces on near-Earth asteroids.\ Nature 
463, 331-334. 

\bibitem[Bottke et al.(1993)]{bot93} Bottke, W.~F., Jr., 
Nolan, M.~C., Greenberg, R.\ 1993.\ Collision lifetimes and impact 
statistics of near-Earth asteroids.\ Lunar and Planetary Science Conference 
24, 159-160. 

\bibitem[Bottke et al.(1994)]{bot94} Bottke, W.~F., Nolan, 
M.~C., Greenberg, R., Kolvoord, R.~A.\ 1994.\ Velocity distributions among 
colliding asteroids.\ Icarus 107, 255-268. 

\bibitem[Bottke et al.(2002)]{bot02} Bottke, W.~F., 
Morbidelli, A., Jedicke, R., Petit, J.-M., Levison, H.~F., Michel, P., 
Metcalfe, T.~S.\ 2002.\ Debiased Orbital and Absolute Magnitude 
Distribution of the Near-Earth Objects.\ Icarus 156, 399-433. 




\bibitem[Bottke et al.(2005)]{bot05a} Bottke, W.~F., Durda, 
D.~D., Nesvorn{\'y}, D., Jedicke, R., Morbidelli, A., Vokrouhlick{\'y}, D., 
Levison, H.\ 2005.\ The fossilized size distribution of the main asteroid 
belt.\ Icarus 175, 111-140. 


\bibitem[Bottke et al.(2005)]{bot05b} Bottke, W.~F., Durda, 
D.~D., Nesvorn{\'y}, D., Jedicke, R., Morbidelli, A., Vokrouhlick{\'y}, D., 
Levison, H.~F.\ 2005.\ Linking the collisional history of the main asteroid 
belt to its dynamical excitation and depletion.\ Icarus 179, 63-94. 




\bibitem[Bowell et al.(1978)]{bow78} Bowell, E., Chapman, 
C.~R., Gradie, J.~C., Morrison, D., Zellner, B.\ 1978.\ Taxonomy of 
asteroids.\ Icarus 35, 313-335. 


\bibitem[Bowell and Lumme(1979)]{bow79} Bowell, E., Lumme, 
K.\ 1979.\ Colorimetry and magnitudes of asteroids.\ Asteroids 132-169. 


\bibitem[Brunetto et al.(2006)]{bru06} Brunetto, R., 
Vernazza, P., Marchi, S., Birlan, M., Fulchignoni, M., Orofino, V., 
Strazzulla, G.\ 2006.\ Modeling asteroid surfaces from observations and 
irradiation experiments: The case of 832 Karin.\ Icarus 184, 327-337. 

%\bibitem[Brunetto et al.(2014)]{bru14} Brunetto, R., and 15 
%colleagues 2014.\ Ion irradiation of Allende meteorite probed by visible, 
%IR, and Raman spectroscopies.\ Icarus 237, 278-292. 


\bibitem[Bus and Binzel(2002a)]{bus02a} Bus, S.~J., Binzel, 
R.~P.\ 2002.\ Phase II of the Small Main-Belt Asteroid Spectroscopic 
Survey. The Observations.\ Icarus 158, 106-145. 


\bibitem[Bus and Binzel(2002b)]{bus02b} Bus, S.~J., Binzel, 
R.~P.\ 2002.\ Phase II of the Small Main-Belt Asteroid Spectroscopic 
Survey. A Feature-Based Taxonomy.\ Icarus 158, 146-177. 

\bibitem[Carvano et 
al.(2010)]{car10} Carvano, J.~M., Hasselmann, P.~H., Lazzaro, D., Moth{\'e}-Diniz, T.\ 2010.\ SDSS-based taxonomic classification and orbital distribution of main belt asteroids.\ Astronomy and Astrophysics 510, AA43. 

\bibitem[Chang et al.(2014)]{cha14} Chang, C.-K., Waszczak, 
A., Lin, H.-W., Ip, W.-H., Prince, T.~A., Kulkarni, S.~R., Laher, R., 
Surace, J.\ 2014.\ A New Large Super-fast Rotator: (335433) 2005 UW163.\ 
The Astrophysical Journal 791, L35. 

\bibitem[Chapman(1996)]{cha96} Chapman, C.~R.\ 1996.\ S-Type Asteroids, Ordinary Chondrites, and Space Weathering: The Evidence from Galileo's Fly-bys of Gaspra and Ida.\ Meteoritics and Planetary Science 31, 699-725. 

\bibitem[Chapman(2004)]{cha04} Chapman, C.~R.\ 2004.\ Space 
Weathering of Asteroid Surfaces.\ Annual Review of Earth and Planetary 
Sciences 32, 539-567. 

\bibitem[Chapman(2010)]{cha10} Chapman, C.~R.\ 2010.\ 
Asteroids: Stripped on passing by Earth.\ Nature 463, 305-306. 



\bibitem[Clark et 
al.(2001)]{cla01} Clark, B.~E., and 11 colleagues 2001.\ Space weathering on Eros: Constraints from albedo and spectral measurements of Psyche crater.\ Meteoritics and Planetary Science 36, 1617-1637. 


\bibitem[Clark et al.(2002)]{cla02} Clark, B.~E., 
Helfenstein, P., Bell, J.~F., Peterson, C., Veverka, J., Izenberg, N.~I., 
Domingue, D., Wellnitz, D., McFadden, L.\ 2002.\ NEAR Infrared Spectrometer 
Photometry of Asteroid 433 Eros.\ Icarus 155, 189-204. 


\bibitem[Dandy et al.(2003)]{dan03} Dandy, C.~L., 
Fitzsimmons, A., Collander-Brown, S.~J.\ 2003.\ Optical colors of 56 
near-Earth objects: trends with size and orbit.\ Icarus 163, 363-373. 

\bibitem[Davis et al.(2002)]{dav02} Davis, D.~R., Durda, 
D.~D., Marzari, F., Campo Bagatin, A., Gil-Hutton, R.\ 2002.\ Collisional 
Evolution of Small-Body Populations.\ Asteroids III 545-558. 

\bibitem[Delbo et al.(2014)]{del14} Delbo, M., Libourel, G., 
Wilkerson, J., Murdoch, N., Michel, P., Ramesh, K.~T., Ganino, C., Verati, 
C., Marchi, S.\ 2014.\ Thermal fatigue as the origin of regolith on small 
asteroids.\ Nature 508, 233-236. 


\bibitem[DeMeo et al.(2009)]{dem09} DeMeo, F.~E., Binzel, 
R.~P., Slivan, S.~M., Bus, S.~J.\ 2009.\ An extension of the Bus asteroid 
taxonomy into the near-infrared.\ Icarus 202, 160-180. 

\bibitem[DeMeo and Carry(2013)]{dem13} DeMeo, F.~E., Carry, 
B.\ 2013.\ The taxonomic distribution of asteroids from multi-filter 
all-sky photometric surveys.\ Icarus 226, 723-741. 

\bibitem[DeMeo and Carry(2014)]{dem14a} DeMeo, F.~E., Carry, 
B.\ 2014.\ Solar System evolution from compositional mapping of the 
asteroid belt.\ Nature 505, 629-634. 


\bibitem[DeMeo et al.(2014)]{dem14b} DeMeo, F.~E., Binzel, 
R.~P., Lockhart, M.\ 2014.\ Mars encounters cause fresh surfaces on some 
near-Earth asteroids.\ Icarus 227, 112-122. 





\bibitem[Hapke(2001)]{hap01} Hapke, B.\ 2001.\ Space 
weathering from Mercury to the asteroid belt.\ Journal of Geophysical 
Research 106, 10039-10074. 

%\bibitem[Hiroi et al.(2013)]{hir13} Hiroi, T., Sasaki, S., 
%Misu, T., Nakamura, T.\ 2013.\ Keys to Detect Space Weathering on Vesta: 
%Changes of Visible and Near-Infrared Reflectance Spectra of HEDs and 
%Carbonaceous Chondrites.\ Lunar and Planetary Science Conference 44, 1276. 


\bibitem[Holmberg et al.(2006)]{hol06} Holmberg, J., Flynn, 
C., Portinari, L.\ 2006.\ The colours of the Sun.\ Monthly Notices of the 
Royal Astronomical Society 367, 449-453.

\bibitem[Holsapple(2007)]{hol07} Holsapple, K.~A.\ 2007.\ 
Spin limits of Solar System bodies: From the small fast-rotators to 2003 
EL61.\ Icarus 187, 500-509. 


\bibitem[Ishiguro et 
al.(2007)]{ish07} Ishiguro, M., and 18 colleagues 2007.\ Global mapping of the degree of space weathering on asteroid 25143 Itokawa by Hayabusa/AMICA observations.\ Meteoritics and Planetary Science 42, 1791-1800. 


\bibitem[Ivezi{\'c} et al.(2001)]{ive01} Ivezi{\'c}, {\v Z}., 
and 32 colleagues 2001.\ Solar System Objects Observed in the Sloan Digital 
Sky Survey Commissioning Data.\ The Astronomical Journal 122, 2749-2784. 


\bibitem[Jacobson and Scheeres(2011)]{jac11} Jacobson, S.~A., 
Scheeres, D.~J.\ 2011.\ Dynamics of rotationally fissioned asteroids: 
Source of observed small asteroid systems.\ Icarus 214, 161-178. 


\bibitem[Jedicke et al.(2004)]{jed04} Jedicke, R., 
Nesvorn{\'y}, D., Whiteley, R., Ivezi{\'c}, {\v Z}., Juri{\'c}, M.\ 2004.\ 
An age-colour relationship for main-belt S-complex asteroids.\ Nature 429, 
275-277. 

\bibitem[Lantz et 
al.(2013)]{lan13} Lantz, C., Clark, B.~E., Barucci, M.~A., Lauretta, D.~S.\ 2013.\ Evidence for the effects of space weathering spectral signatures on low albedo asteroids.\ Astronomy and Astrophysics 554, A138. 

\bibitem[Lazzaro et al.(2004)]{laz04} Lazzaro, D., Angeli, 
C.~A., Carvano, J.~M., Moth{\'e}-Diniz, T., Duffard, R., Florczak, M.\ 
2004.\ S $^{3}$OS $^{2}$: the visible spectroscopic survey of 820 
asteroids.\ Icarus 172, 179-220.

%\bibitem[Lazzarin et al.(2006)]{laz06} Lazzarin, M., Marchi, 
%S., Moroz, L.~V., Brunetto, R., Magrin, S., Paolicchi, P., Strazzulla, G.\ 
%2006.\ Space Weathering in the Main Asteroid Belt: The Big Picture.\ The 
%Astrophysical Journal 647, L179-L182. 


\bibitem[Magnier et al.(2013)]{mag13} Magnier, E.~A., and 13 
colleagues 2013.\ The Pan-STARRS 1 Photometric Reference Ladder, Release 
12.01.\ The Astrophysical Journal Supplement Series 205, 20. 


\bibitem[Marchi et al.(2006)]{mar06} Marchi, S., Magrin, S., 
Nesvorn{\'y}, D., Paolicchi, P., Lazzarin, M.\ 2006.\ A spectral slope 
versus perihelion distance correlation for planet-crossing asteroids.\ 
Monthly Notices of the Royal Astronomical Society 368, L39-L42. 


\bibitem[Migliorini et al.(1998)]{mig98} Migliorini, F., 
Michel, P., Morbidelli, A., Nesvorny, D., Zappala, V.\ 1998.\ Origin of 
Multikilometer Earth- and Mars-Crossing Asteroids: A Quantitative 
Simulation.\ Science 281, 2022. 


\bibitem[Miyazaki et al.(2002)]{miy02} Miyazaki, S., and 14 
colleagues 2002.\ Subaru Prime Focus Camera -- Suprime-Cam.\ Publications 
of the Astronomical Society of Japan 54, 833-853. 


\bibitem[Morbidelli and Vokrouhlick{\'y}(2003)]{mor03} 
Morbidelli, A., Vokrouhlick{\'y}, D.\ 2003.\ The Yarkovsky-driven origin of 
near-Earth asteroids.\ Icarus 163, 120-134. 

\bibitem[Morbidelli et al.(2009)]{mor09} Morbidelli, A., 
Bottke, W.~F., Nesvorn{\'y}, D., Levison, H.~F.\ 2009.\ Asteroids were born 
big.\ Icarus 204, 558-573. 


%\bibitem[Moroz et al.(2004)]{mor04} Moroz, L.~V., Hiroi, T., 
%Shingareva, T.~V., Basilevsky, A.~T., Fisenko, A.~V., Semjonova, L.~F., 
%Pieters, C.~M.\ 2004.\ Reflectance Spectra of CM2 Chondrite Mighei 
%Irradiated with Pulsed Laser and Implications for Low-Albedo Asteroids and 
%Martian Moons.\ Lunar and Planetary Science Conference 35, 1279. 

\bibitem[Moth{\'e}-Diniz and 
Nesvorn{\'y}(2008)]{mot08} Moth{\'e}-Diniz, T., Nesvorn{\'y}, D.\ 2008.\ Visible spectroscopy of extremely young asteroid families.\ Astronomy and Astrophysics 486, L9-L12.


\bibitem[Nakamura and Yoshida(2002)]{nak02} Nakamura, T., 
Yoshida, F.\ 2002.\ Statistical Method for Deriving Spatial and Size 
Distributions of Sub-km Main-Belt Asteroids from Their Sky Motions.\ 
Publications of the Astronomical Society of Japan 54, 1079-1089. 

\bibitem[Nesvorn{\'y} et al.(2005)]{nes05} Nesvorn{\'y}, D., 
Jedicke, R., Whiteley, R.~J., Ivezi{\'c}, {\v Z}.\ 2005.\ Evidence for 
asteroid space weathering from the Sloan Digital Sky Survey.\ Icarus 173, 
132-152. 

\bibitem[Nesvorn{\'y} et al.(2010)]{nes10} Nesvorn{\'y}, D., 
Bottke, W.~F., Vokrouhlick{\'y}, D., Chapman, C.~R., Rafkin, S.\ 2010.\ Do 
planetary encounters reset surfaces of near Earth asteroids?.\ Icarus 209, 
510-519. 





\bibitem[Padmanabhan et al.(2008)]{pad08} Padmanabhan, N., 
and 23 colleagues 2008.\ An Improved Photometric Calibration of the Sloan 
Digital Sky Survey Imaging Data.\ The Astrophysical Journal 674, 1217-1233. 


\bibitem[Petit et al.(2006)]{pet06} Petit, J.-M., Holman, 
M.~J., Gladman, B.~J., Kavelaars, J.~J., Scholl, H., Loredo, T.~J.\ 2006.\ 
The Kuiper Belt luminosity function from m$_{R}$= 22 to 25.\ Monthly 
Notices of the Royal Astronomical Society 365, 429-438. 


\bibitem[Polishook et al.(2014)]{pol14} Polishook, D., 
Moskovitz, N., Binzel, R.~P., DeMeo, F.~E., Vokrouhlick{\'y}, D., {\v 
Z}i{\v z}ka, J., Oszkiewicz, D.\ 2014.\ Observations of 
fresh and weathered surfaces on asteroid pairs and their 
implications on the rotational-fission mechanism.\ Icarus 233, 9-26. 


\bibitem[Rabinowitz(1997)]{rab97} Rabinowitz, D.~L.\ 1997.\ 
Are Main-Belt Asteroids a Sufficient Source for the Earth-Approaching 
Asteroids?.\ Icarus 127, 33-54. 

\bibitem[Rivkin et al.(2011)]{riv11} Rivkin, A.~S., Thomas, 
C.~A., Trilling, D.~E., Enga, M.-t., Grier, J.~A.\ 2011.\ Ordinary 
chondrite-like colors in small Koronis family members.\ Icarus 211, 
1294-1297. 

\bibitem[Rozitis et al.(2014)]{roz14} Rozitis, B., Maclennan, 
E., Emery, J.~P.\ 2014.\ Cohesive forces prevent the rotational breakup of 
rubble-pile asteroid (29075) 1950 DA.\ Nature 512, 174-176. 

\bibitem[Rubincam(2000)]{rub00} Rubincam, D.~P.\ 2000.\ 
Radiative Spin-up and Spin-down of Small Asteroids.\ Icarus 148, 2-11.

\bibitem[Sasaki et al.(2001)]{sas01} Sasaki, S., Nakamura, 
K., Hamabe, Y., Kurahashi, E., Hiroi, T.\ 2001.\ Production of iron 
nanoparticles by laser irradiation in a simulation of lunar-like space 
weathering.\ Nature 410, 555-557. 


\bibitem[Schlafly et al.(2012)]{sch12} Schlafly, E.~F., and 
16 colleagues 2012.\ Photometric Calibration of the First 1.5 Years of the 
Pan-STARRS1 Survey.\ The Astrophysical Journal 756, 158. 


\bibitem[Scheeres(2007)]{sch07} Scheeres, D.~J.\ 2007.\ The 
dynamical evolution of uniformly rotating asteroids subject to YORP.\ 
Icarus 188, 430-450. 

\bibitem[Scheeres(2015)]{sch15} Scheeres, D.~J.\ 2015.\ 
Landslides and Mass shedding on spinning spheroidal asteroids.\ Icarus 247, 
1-17. 


\bibitem[Stuart and Binzel(2004)]{stu04} Stuart, J.~S., 
Binzel, R.~P.\ 2004.\ Bias-corrected population, size distribution, and 
impact hazard for the near-Earth objects.\ Icarus 170, 295-311. 


\bibitem[Tholen(1984)]{tho84} Tholen, D.~J.\ 1984.\ Asteroid 
taxonomy from cluster analysis of Photometry.\ Ph.D.~Thesis . 

\bibitem[Thomas et al.(2011)]{tho11} Thomas, C.~A., Rivkin, 
A.~S., Trilling, D.~E., Enga, M.-t., Grier, J.~A.\ 2011.\ Space weathering 
of small Koronis family members.\ Icarus 212, 158-166.

\bibitem[Thomas et al.(2012)]{tho12} Thomas, C.~A., Trilling, 
D.~E., Rivkin, A.~S.\ 2012.\ Space weathering of small Koronis family 
asteroids in the SDSS Moving Object Catalog.\ Icarus 219, 505-507. 


\bibitem[Tonry et al.(2012)]{ton12} Tonry, J.~L., and 13 
colleagues 2012.\ The Pan-STARRS1 Photometric System.\ The Astrophysical 
Journal 750, 99. 


\bibitem[Vere{\v s} et al.(2012)]{ver12} Vere{\v s}, P., 
Jedicke, R., Denneau, L., Wainscoat, R., Holman, M.~J., Lin, H.-W.\ 2012.\ 
Improved Asteroid Astrometry and Photometry with Trail Fitting.\ 
Publications of the Astronomical Society of the Pacific 124, 1197-1207. 


\bibitem[Vernazza et al.(2009)]{ver09} Vernazza, P., Binzel, 
R.~P., Rossi, A., Fulchignoni, M., Birlan, M.\ 2009.\ Solar wind as the 
origin of rapid reddening of asteroid surfaces.\ Nature 458, 993-995. 

\bibitem[Walsh et al.(2012)]{wal12} Walsh, K.~J., Richardson, 
D.~C., Michel, P.\ 2012.\ Spin-up of rubble-pile asteroids: Disruption, 
satellite formation, and equilibrium shapes.\ Icarus 220, 514-529. 

\bibitem[Willman et al.(2008)]{wil08} Willman, M., Jedicke, 
R., Nesvorn{\'y}, D., Moskovitz, N., Ivezi{\'c}, {\v Z}., Fevig, R.\ 2008.\ 
Redetermination of the space weathering rate using spectra of Iannini 
asteroid family members.\ Icarus 195, 663-673. 

\bibitem[Willman et al.(2010)]{wil10} Willman, M., Jedicke, 
R., Moskovitz, N., Nesvorn{\'y}, D., Vokrouhlick{\'y}, D., Moth{\'e}-Diniz, 
T.\ 2010.\ Using the youngest asteroid clusters to constrain the space 
weathering and gardening rate on S-complex asteroids.\ Icarus 208, 758-772. 

\bibitem[Willman and Jedicke(2011)]{wil11} Willman, M., 
Jedicke, R.\ 2011.\ Asteroid age distributions determined by space 
weathering and collisional evolution models.\ Icarus 211, 504-510. 


\bibitem[Yoshida and 
Nakamura(2007)]{yos07} Yoshida, F., Nakamura, T.\ 2007.\ Subaru Main Belt Asteroid Survey (SMBAS) - Size and color distributions of small main-belt asteroids.\ Planetary and Space Science 55, 1113-1125. 


\bibitem[Zellner et al.(1985)]{zel85} Zellner, B., Tholen, 
D.~J., Tedesco, E.~F.\ 1985.\ The eight-color asteroid survey - Results for 
589 minor planets.\ Icarus 61, 355-416.



\end{thebibliography}
\end{document}